\documentclass[aps,prd, two column]{revtex4-2}

\usepackage{amsmath}
\usepackage{ulem}
\usepackage{graphicx}
\usepackage{epsfig}
\usepackage{bm}
\usepackage{natbib}
\usepackage{pgfplots,mathtools}
\usepackage{hyperref}
\usepackage{amsmath}
\usepackage{braket}
\usepackage{slashed}
\usepackage[compat=1.0.0]{tikz-feynman}
\usepackage{physics}
\usepackage{caption}
\usepackage{subcaption}
\usepackage{xfrac}
\usepackage{algorithm}

\newcommand{\bef}{\begin{figure}}
\newcommand{\eef}{\end{figure}}
\newcommand{\bc}{\begin{center}}
\newcommand{\ec}{\end{center}}
\newcommand{\nn}{\nonumber}
\newcommand{\be}{\begin{equation}}
\newcommand{\ee}{\end{equation}}
\newcommand{\bea}{\begin{eqnarray}}
\newcommand{\eea}{\end{eqnarray}}

\def\ba{\begin{eqnarray}}
\def\ea{\end{eqnarray}}
\newcommand{\ep}{\epsilon}
\newcommand{\om}{\omega} 
\newcommand{\Om}{\Omega}

\newcommand{\del}{\partial}
\newcommand{\al}{\alpha}
\newcommand{\bt}{\beta}
\newcommand{\la}{\lambda}

\newcommand{\orcid}[1]{\href{https://orcid.org/#1}{\includegraphics[width=8pt]
{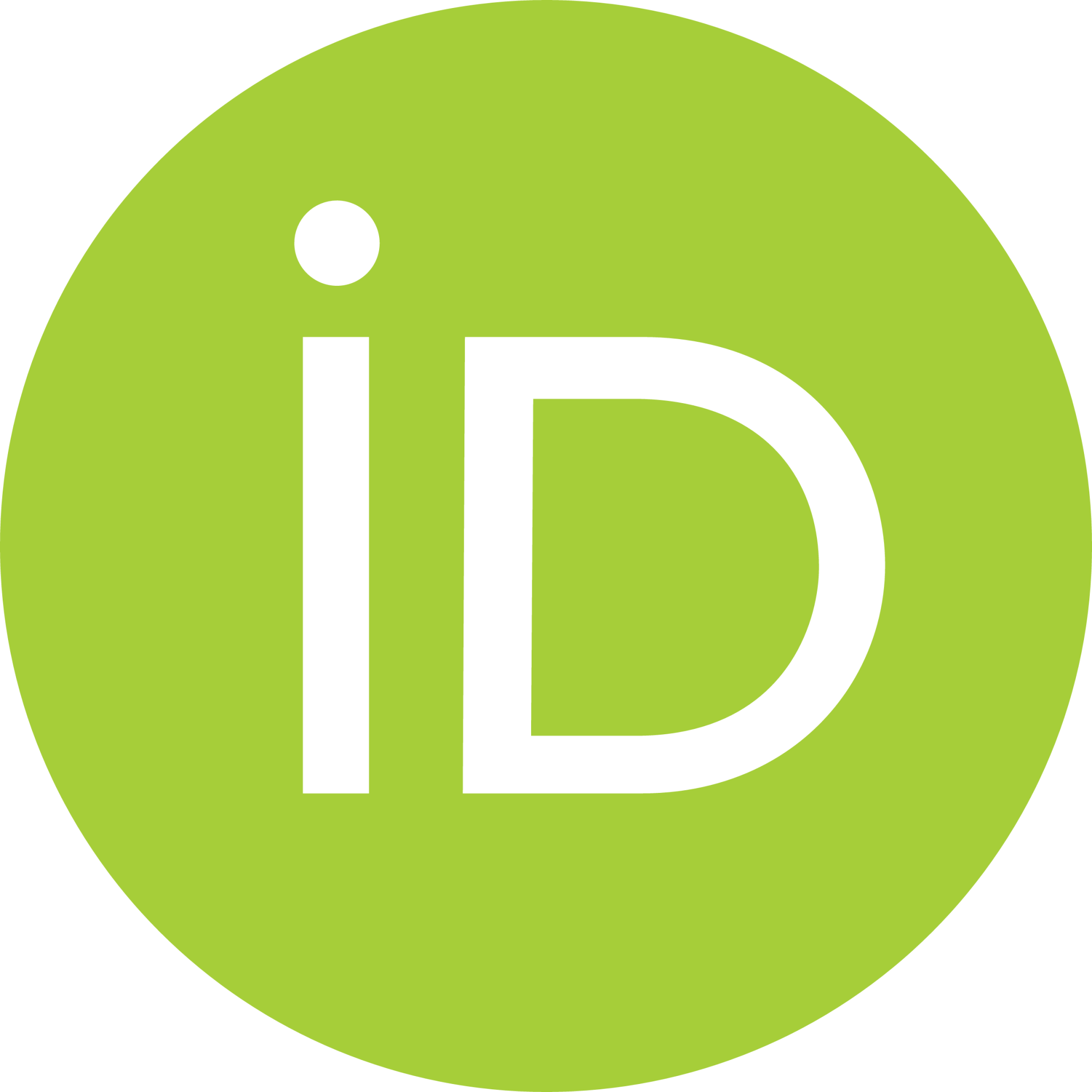}}}
\begin{document}
\title{Shear viscosity and electrical conductivity of rotating quark matter in Nambu-Jona Lasinio Model}

\author{Ashutosh Dwibedi\orcid{0009-0004-1568-2806}$^1$, 
Dushmanta Sahu\orcid{0000-0001-8980-1362}$^{2}$, 
Jayanta Dey\orcid{0000-0002-0894-6402}$^{1, 3}$, 
Kangkan Goswami\orcid{0000-0002-0476-1005}$^{4}$}
\author{Sabyasachi Ghosh\orcid{0000-0003-1212-824X}$^1$, Raghunath Sahoo\orcid{0000-0003-3334-0661}$^{4}$}
\affiliation{$^{1}$Department of Physics, Indian Institute of Technology Bhilai, Kutelabhata, Durg 491001, India}
\affiliation{$^{2}$ Instituto de Ciencias Nucleares, Universidad Nacional Autónoma de México, Apartado Postal 70-543,
México Distrito Federal 04510, México}
\affiliation{$^{3}$ Bogoliubov Laboratory of Theoretical Physics, Joint Institute for Nuclear Research, Dubna, 141980, Russia,}
\affiliation{$^{4}$Department of Physics, Indian Institute of Technology Indore, Simrol, Indore 453552, India}

\begin{abstract}
    The Lagrangian for strongly interacting and rotating quark matter is modified with the inclusion of the spinorial connections, which in turn affect the thermodynamic equation of state and transport properties of the medium.
    In this work, we investigate the transport properties of quark matter under finite rotation, focusing specifically on electrical conductivity and shear viscosity by using a two-flavor Nambu-Jona-Lasinio (NJL) model. The chiral condensate in the NJL model decreases under rotation, leading to enhanced transport properties. Moreover, rotation induces anisotropy in the transport coefficients, which are calculated within the kinetic theory framework using the Boltzmann transport equation. The Coriolis force is introduced in the force term of the Boltzmann transport equation, like the Lorentz force, which is considered for finite magnetic fields. By using a phenomenological temperature-dependent angular velocity, we observe that the variation of anisotropic components with temperature preserves the traditional valley-shaped pattern. However, the magnitude of the anisotropic components is suppressed compared to the usual component one finds in the absence of rotation. Interestingly, at zero net quark density, Hall-like transport phenomena emerge as significant non-dissipative contributions under rotation, which is not expected under finite magnetic fields due to the cancellation of quark and anti-quark Hall currents.

\end{abstract}
    
\maketitle
\section{Introduction}

One of the main goals of the Large Hadron Collider (LHC) and Relativistic Heavy Ion Collider (RHIC) is to understand the hot and dense deconfined matter, which was supposed to be prevalent in the universe’s infancy. This deconfined matter is locally thermalized and is called the quark-gluon plasma (QGP)~\cite{Shuryak:2003xe,Heinz:2004qz}. The temperature at which the hadrons melt to produce QGP is about $10^5$ times that of the Sun’s core. In laboratories, these conditions are achieved by colliding ultra-relativistic nuclei that produce QGP, which is short-lived. At such high temperatures and extreme densities, the matter behaves very uniquely. Firstly, due to the fast motion of charged spectators, a large transient electromagnetic field is created in peripheral heavy-ion collisions. The strength of this magnetic field ($eB$) is around $m_\pi^{2} \sim 10^{18}$ G  at RHIC to  $15m_\pi^2$ at LHC~\cite{Skokov:2009qp}. In comparison, the strongest magnetic field created on Earth in the form of shock waves is around $10^7$ G, and the magnetic field created in the neutron stars is around $10^{10} - 10^{13}$ G. Thus, strong magnetic field in the heavy-ion collisions affect the equation of state of the created matter~\cite{Bzdak:2011yy}.

In addition to a strong magnetic field, a huge initial angular momentum is produced in peripheral heavy-ion collisions. The angular momentum is then distributed to the particles that are produced in the collisions. The orbital angular momentum of the particles gets coupled with their spins in a thermalized system, which can later be observed as the polarization of the particles. Evidence of this was found recently when the STAR experiment showed a finite hyperon polarization in heavy-ion collisions~\cite{STAR:2022fan}. Rotation or vorticity can also be generated from the Einstein-de Haas effect, where a magnetized medium creates a finite rotation~\cite{1915DPhyG17152E,1908Richardson}. This suggests that the substantial magnetic field produced in heavy ion collisions due to fast-moving spectators may magnetize the medium and generate a large amount of vorticity in the system. The reverse effect is called the Barnett effect. An analogy between rotation and the magnetic field is a well-known phenomenon studied in many physical systems~\cite{J_Sivardiere_1983,Johnson2000-px,Sakurai1980,BRANDAO201555,Chen:2015hfc,Mameda:2015ria}. Much like the external magnetic field playing a crucial role in the Equation of State (EoS) of the system, angular velocity ($\Omega$) can also be a key player that can affect the EoS. While the effect of the magnetic field on the thermodynamic and transport properties of the hot and dense partonic and hadronic matter has been studied extensively~\cite{Ghosh:2019ubc,Dey:2020awu,Kalikotay:2020snc,Dey:2021fbo,Satapathy:2021cjp,Das:2019wjg,Das:2019ppb,Chatterjee:2019nld,Hattori:2016cnt,Hattori:2016lqx,Satapathy:2021wex,Andersen:2021lnk,Li:2020dwr, Goswami:2023eol,Sahoo:2023vkw}, the effect of rotation has been relatively less explored~\cite{Becattini:2021lfq,Wang:2018sur,Fujimoto:2021xix,Pradhan:2023rvf}.

One of the robust methods to understand the initial dynamics of the deconfined matter is the first-principle-based lattice QCD (lQCD). However, at non-vanishing baryochemical potential ($\mu_{B}$), the numerical simulation has to deal with the fermion sign problem, which does not have an exact solution. Recently, the effect of rotation has been explored with lQCD, where the authors have estimated how the real and imaginary angular momentum affects the Polyakov loop~\cite{Braguta:2021jgn}. Regardless, other interesting and highly effective models have been used to study the hot and dense matter, which agree with the lQCD predictions and can also work in baryon-rich environments. The Nambu-Jona-Laisino (NJL) model is one such alternative~\cite{Nambu:1961fr,Nambu:1961tp,Hatsuda:1994pi,Klevansky:1992qe}. It is an effective field theoretical model that encapsulates important aspects of quantum chromodynamics, specifically spontaneous chiral symmetry breaking and dynamical mass creation. The NJL model allows the study of quark condensates and their role in producing constituent quark masses by substituting an efficient four-fermion contact interaction for gluonic interactions. Since the NJL framework has been widely used to study the QCD phase diagram, including phase transitions at finite temperature and density, it is a useful tool for examining the characteristics of quark matter under extreme circumstances, like those present in compact astrophysical objects and heavy-ion collisions. In this work, we use the 2-flavor NJL model under rotation and estimate the electrical conductivity ($\sigma$) and shear viscosity ($\eta$) of the system. When studying heavy-ion collisions, the shear viscosity to entropy density ratio ($\eta/s$) and electrical conductivity are essential transport coefficients because they reveal information about the characteristics of the QGP and how it changes over time. The ratio $\eta/s$ is a measure of the fluidity of the system, where a lower value indicates a more perfect fluid behavior. The QGP formed in ultrarelativistic heavy-ion collisions at RHIC and LHC is thought to have a near-minimal $\eta/s$, which approaches the Kovtun-Son-Starinets (KSS) bound $1/4\pi$~\cite{Kovtun:2004de}.  This small $\eta/s$ affects the QGP’s thermalization and hydrodynamic response and is crucial to collective flow phenomena like elliptic and triangular flow. Conversely, the electrical conductivity controls how the QGP reacts to external electromagnetic fields and influences the creation of phenomena like electromagnetic emissions and the chiral magnetic effect.  Determining $\sigma$ precisely aids in comprehending how the QGP interacts with strong magnetic fields created in non-central heavy-ion collisions. $\sigma$ primarily governs the response of the plasma to induced electric fields and plays a crucial role in electromagnetic radiation processes. Due to their sensitivity to temperature, chemical potential, and angular velocity, $\eta/s$ and $\sigma$ are both crucial for investigating the QCD phase diagram and the transition between hadronic and partonic phases. To evaluate the transport coefficients, we have used the Boltzmann transport equation (BTE) in the standard relaxation time approximation (RTA). The rotational background in the BTE has been included through the connection coefficients, which are related to the derivative of the rotating frame metric. In this way, naturally, the apparent forces, i.e., Coriolis and Centrifugal, come into existence, which make the system's transport properties anisotropic. The transport coefficient will have distinct components, which we have studied as a function of temperature and angular velocity.

Recent lattice QCD studies~\cite{Yang:2023vsw, Braguta:2024zpi} found contradictory results from those of effective models. These first-principle studies show that the chiral condensate is enhanced and the Polyakov loop is reduced due to real rotation. In other words, catalysis of condensate and confinement occur due to real rotation both in pure SU(3) gluon dynamics and staggered fermion lQCD. Very recently, the NJL model-based study~\cite{Jiang2022} by Yin Jiang shows that the $\Omega$ dependent running coupling $G(\Omega)$ reverses the trend of condensate to justify effective gluon coupling. A recent work~\cite{Nunes:2024hzy} considered the $T$ and $\Omega$ dependent coupling constant ($G(T, \Omega)$) constrained by the pseudo-critical temperature obtained in lQCD simulation at finite rotation. However, in the current work, we have studied the transport properties of rotating quark matter for the first time using the NJL model, and we have considered a fixed coupling constant for the same. 

The paper is organized as follows: in Sec.~\ref{NJL}, we briefly describe the NJL Lagrangian in the rotating frame and write down the grand potential energy and the subsequent gap equation for the constituent quark mass. Thereafter, we describe the quasi-particle BTE for the quarks and write down the expressions of electrical conductivities and shear viscosities in Sec.~\ref{BTE}. In Sec.~\ref{sec-result}, we show our results for the variation of anisotropic viscosities and conductivities as a function of angular velocity and temperature. Subsequently, we also take a phenomenological temperature-dependent angular velocity to see the realistic variation of transport coefficients with temperature.
In the end, we summarize the article and make concluding remarks in Sec.~\ref{Section-Summary}.
\section{Formulation}
\label{formu}
\subsection{NJL model in rotating frame}\label{NJL}

The Lagrangian for the two-flavor NJL model in a rotating frame is given by~\cite{Jiang:2016wvv,Chen:2015hfc,Chernodub_2017,Ebihara:2016fwa,Wang:2018sur,Chernodub:2017ref},
\bea
\mathcal{L} = \bar{\psi}[i\gamma^{\mu}(\partial_{\mu}+\Gamma_{\mu})-m]\psi+G_s[(\bar{\psi}\psi)^2+(\bar{\psi}i\gamma_5\vec{\sigma}\psi)^2],\label{A1}
\eea
where the quark fields $\psi\equiv
\begin{pmatrix}
  \psi_{u} \\
  \psi_{d} \\
\end{pmatrix}$. The spinorial affine connection $\Gamma_{\mu}$ can be expressed as $\Gamma_{\mu}=\frac{1}{8}\omega_{\mu ab} [\gamma^{a},\gamma^{b}]$. The spin connections $\omega_{\mu ab}$ is written in terms of the metric $g^{\mu\nu}$, vierbein fields (tetrads) $e^{\mu}_{~a}$ and the affine connections (Christoffel symbols) $\Gamma^{\mu}_{\al\bt}$ as~\cite{Pollock:2010zz,Yepez:2011bw,Kapusta:2019sad,RevModPhys.29.465}, $\om_{\mu ab}=g_{\nu\bt}e^{\nu}_{~a}(\del_{\mu}e^{\bt}_{~b}+e^{\la}_{~b}\Gamma^{\bt}_{\mu\la})$. The Greek and Latin indices are used for the co-rotating coordinate space ($\mu=t,x,y,z$) and the inertial tangent space $(a=0,1,2,3)$ defined by the vierbein fields, respectively. The vierbein which define the local tangent space is chosen as~\cite{Kapusta:2019sad,Ebihara:2016fwa}, $e^{t}_{~0}=e^{x}_{~1}=e^{y}_{~2}=e^{z}_{~3}=1,~ e^{x}_{~0}=\Om y,~ e^{y}_{~0}=-\Om x$ and zero otherwise. The rotation about the $z$-axis with angular velocity $\vec{\Om}=\Om \hat{k}$ is implemented by the metric $g_{\mu\nu}$ which connects the fixed inertial coordinates with the co-rotating coordinates~\cite{Padhan:2024edf},
\bea
g_{\mu\nu}=
\begin{pmatrix}
      1-\Om^2x^{2}-\Om^2y^{2}  & \Om y & -\Om x & 0\\
    \Om y                           &      -1        &         0       & 0\\
   -\Om x                           &       0        &        -1       & 0 \\
    0                                        &       0        &         0       & -1 
    \label{R2}
\end{pmatrix}.
\eea
The Lagrangian (\ref{A1}) in the mean-field approximation for isospin symmetric matter can be written as~\cite{Wang:2018sur,Chen2021},
\bea
\mathcal{L}=\bar{\psi}[i\gamma^{\mu}(\partial_{\mu}+\Gamma_{\mu})-M]\psi-\frac{(M-m)^{2}}{4G_{S}},\label{A2}
\eea
where the constituent quark mass $M\equiv m-2G_{S}\langle\bar{\psi}\psi \rangle$. The constituent quark mass $M$ is the dynamically generated mass that comes into play as a result of quark self-energy~\cite{Buballa:2003qv}. Starting from the mean-field approximated Lagrangian (\ref{A2}) and using the standard thermal field theory techniques; one obtains the following grand potential energy $\Tilde{\Om}$ for the rotating system of quarks in the local density approximation~\cite{Wang:2018sur,Chen2021},
\begin{widetext}
    \bea
    \Tilde{\Om}&=&\int d^{3}\vec{x}\bigg[\frac{(M-m)^{2}}{4G_{S}}-\frac{N_{c}N_{f}}{8\pi^{2}}\sum_{m=-\infty}^{\infty}\int_{0}^{\Lambda} dp_{\perp}^{2} \int_{-\sqrt{\Lambda^{2}-p_{\perp}^{2}}}^{\sqrt{\Lambda^{2}-p_{\perp}^{2}}} dp_{z} ~ E \left(J^{2}_{m}(p_{\perp}\rho)+J^{2}_{m+1}(p_{\perp}\rho)\right)\nn\\
    &-&\frac{TN_{c}N_{f}}{8\pi^{2}}\sum_{m=-\infty}^{\infty}\int dp_{\perp}^{2} \int dp_{z}\left(\ln(1+e^{-\bt(E-(m+\frac{1}{2})\Om)}) +\ln (1+e^{-\bt(E+(m+\frac{1}{2})\Om)})\right)\left(J^{2}_{m}(p_{\perp}\rho)+J^{2}_{m+1}(p_{\perp}\rho)\right)\bigg],\nonumber\\
    \label{A3}
    \eea
     where $J_{m}$ is the $m$th order Bessel function of the first kind and $p_{\perp}\equiv \sqrt{p_{x}^{2}+p_{y}^{2}}$, $\rho\equiv\sqrt{x^{2}+y^{2}}$ are the radial momentum and coordinate respectively. The particle energy $\Tilde{E}$ in a co-rotating frame can be expressed as~\cite{Ambru__2014,Ambru__2016}, $\Tilde{E}=E-(m+\frac{1}{2})\Om$, where $E=\sqrt{p_{\perp}^{2}+p_{z}^{2}+M^{2}}$. Since the NJL model is not renormalizable, one needs to prescribe a scheme to regularize the divergences appearing in the vacuum part of Eq.~(\ref{A3}). We here adhere to regularize the integrals by introducing the 3-momentum cut-off $\Lambda$. The mass gap equation is obtained by minimizing $\Tilde{\Om}$ in Eq.~(\ref{A3}) with respect to the constituent mass $M$ at each space point~\cite{Wang:2018sur}, i.e.,  
    \bea
    && \frac{\del \Tilde{\Om}}{\del M}=0,\label{A4}\\
    \text{with the constraint, } &&\frac{\del^{2} \Tilde{\Om}}{\del M^{2}}> 0.\label{A5}
    \eea
    Using Eq.~(\ref{A4}), one obtains the following explicit form of the gap equation for the constituent mass $M$~\cite{Chen2021},
    \bea
    M=m&+&\frac{G_{S}N_{c}N_{f}}{4\pi^{2}}\sum_{m=-\infty}^{\infty}\int_{0}^{\Lambda} dp_{\perp}^{2} \int_{-\sqrt{\Lambda^{2}-p_{\perp}^{2}}}^{\sqrt{\Lambda^{2}-p_{\perp}^{2}}} dp_{z}\left(J^{2}_{m}(p_{\perp}\rho)+J^{2}_{m+1}(p_{\perp}\rho)\right)\frac{M}{E}\nn\\
    &-&\frac{G_{S}N_{c}N_{f}}{4\pi^{2}}\sum_{m=-\infty}^{\infty}\int dp_{\perp}^{2} \int dp_{z}\left(J^{2}_{m}(p_{\perp}\rho)+J^{2}_{m+1}(p_{\perp}\rho)\right)\frac{M}{E}\left[\frac{1}{e^{\bt(E-(m+\frac{1}{2})\Om)}+1}+\frac{1}{e^{\bt(E+(m+\frac{1}{2})\Om)}+1}\right].\nonumber\\
    \label{A6}
    \eea
\end{widetext}
One has to solve Eq.~(\ref{A6}) along with the constraint~(\ref{A5}) to get the constituent quark mass as a function of angular velocity ($\Omega$), radius ($\rho$), temperature ($T$), 3-momentum cut-off ($\Lambda$), and the coupling strength ($G_{S}$). However, we can fix the 3-momentum cut-off $\Lambda$ and the coupling strength $G_{S}$ by using the pion mass and decay constant in vacuum~\cite{Buballa:2003qv,Klevansky:1992qe}, so that $M=M(\rho,\Omega,T)$.

Now, we will move on to the next section to evaluate shear viscosity and electrical conductivity for the rotating system of quarks having constituent mass $M(\rho, \Omega, T)$.

\subsection{Shear viscosity and electrical conductivity in rotating frame}\label{BTE}
Here, we will briefly describe the calculation of the transport coefficients under rotation in the kinetic theory formalism. For a more detailed description, see Refs.~\cite{Aung:2023pjf, Dwibedi:2023akm,Padhan:2024edf,Dwibedi:2025bdd}.

%The kinetic theory description of the rotating quark or hadronic medium created in the heavy ion collision is described in the Refs.~\cite{Aung:2023pjf, Dwibedi:2023akm,Padhan:2024edf}, readers can get the physical ideas and a clear picture of the approximations involved therein. Here, we will briefly describe the steps which are essential to writing down the final expressions of shear viscosity and electrical conductivity. 

The micro and macroscopic expressions of electrical current density $J^{i}$ ($i$ runs from $1 \text{ to } 3$) and shear stress tensor $\pi^{ij}$ ($i, j$ run from $1 \text{ to } 3$) for the rotating system of the dressed quarks under an applied electric field $\Vec{\Tilde{E}}\equiv \Tilde{E}\hat{e}$ are,
\begin{align}
   & J^{i} =\sum_{r}J_{r}^{i}= \sum_{r} g_{r} q_{r} \int{\frac{d^{3}\vec{p}_{r}}{{(2\pi})^{3}}} \frac{p^{i}_{r}}{E_r} \delta{f_{r}},\label{A7}\\
&=\sum_{r}\sigma_{r}^{ij}\tilde{E}_{j}\equiv\sigma^{ij}\tilde{E}_{j},\label{A8}\\
 &\pi^{ij}=\sum_{r} \pi^{ij}_{r}=\sum_{r} g_{r} \int\frac{d^{3}\vec{p}_{r}} {{(2\pi})^{3}} \frac{p^{i}_{r}p^{j}_{r}}{E_r} \delta{f_{r}},\label{A9}\\
&=-\sum\limits_{r}\eta^{r}_{n} C^{ij}_{n}\equiv -\sum\limits_{n=0}^{4}\eta_{n} C^{ij}_{n},\label{A10}
\end{align}
where $r\in\{u,\bar{u},d,\bar{d}\}$ is the label characterizing the quarks' charge $q_{r}$, energy $E_r$, momentum $p_r$, and degeneracy $g_{r}=2~(\text{spin})\times 3~(\text{color})=6$. $C^{ij}_{n}$ are five independent traceless tensors containing the fluid velocity gradient  $U^{kl} $=$ \frac{1}{2} (\frac{\partial u_k}{\partial x_l}+\frac{\partial u_l}{\partial x_k})$. In the above, Eqs.~(\ref{A7}) and~(\ref{A9}) define the microscopic evaluation of $J^{i}$ and $\pi^{ij}$ in terms of deviation of constituent quark distributions from local equilibrium $\delta{f_{r}}\equiv f_{r}-f^{0}_{r}$. The local equilibrium distribution for the constituent quarks is given by~\cite{Padhan:2024edf}, $f^{0}_{r}=1/[e^{(u^{\al}p_{r}^{\bt}g_{\al\bt}-\mu_{r})/T}+1]$, where $u^{\mu}=\gamma(1, u^{i})$, $\mu_{r}$ and $p^{\mu}_{r}\equiv (p^{0}_{r},p^{i}_{r})$ are the fluid four-velocity and chemical potential and four-momentum of the quarks, respectively. The macroscopic description of $J^{i}$ and $\pi^{ij}$ laid out in Eqs.~(\ref{A8}) and~(\ref{A10}) are the mathematical expressions of Ohm’s law and Newton’s law of viscosity, respectively. The conductivities and shear viscosities can be determined by explicitly evaluating $J^{i}$ and $\pi^{ij}$ from Eqs.~(\ref{A7}) and~(\ref{A9}) and comparing them with  Eqs.~(\ref{A8}) and~(\ref{A10}). The form of $\delta f_{r}$ needed for the explicit evaluation of the integrals in Eqs.~(\ref{A7}) and~(\ref{A9}) can be obtained with the aid of BTE. The quasi-particle BTE for the rotating quarks with space-time dependent mass can be written as~\cite{Romatschke:2011qp,Tinti:2016bav,Albright:2015fpa,Nogarolli:2024azy,Rocha:2024rce},
\bea
p_{r}^{\mu}~\frac{\del f_{r}}{\del x^{\mu}} &-& p_{r}^{\mu}p_{r}^{\beta}~\Gamma^{\al}_{\mu\bt}~\frac{\del f_{r}}{\del p_{r}^{\al}} + g^{\mu\al}~M \del_{\al}M~ \frac{\del f_{r}}{\del p_{r}^{\mu}} \nn\\
&+& q_{r} F^{\beta\al}p_{r\al}~\frac{\del f_{r}}{\del p_{r}^{\bt}} = C[f_{r}], \label{A11}
\eea
where the second and third terms in the LHS can be interpreted as effective forces, arising because of the rotating space-time geometry and space-time-dependent constituent quark mass $M=M(x^{\mu},\mu(x^{\mu}), T(x^{\mu}))$, respectively. The third term in the LHS is the usual Lorentz force that arises for particles of charge $q$ in the presence of electromagnetic fields represented by the Faraday tensor $F^{\mu\nu}$. The RHS of Eq.~(\ref{A11}) is known as the collision kernel, which arises because of the random momentary collision among the quarks. For solving Eq.~(\ref{A11}) to obtain $\delta f_{r}$, we will assume that the system is close to equilibrium, i.e., $f=f^{0}_{r}+\delta f_{r}$, where $\delta f_{r}$ is a small perturbation and use the well-known relaxation time approximation (RTA). Rewriting Eq.~(\ref{A11}) in the RTA of Anderson-Witting type we have~\cite{ANDERSON1974466,Jaiswal:2013vta},
\begin{align}
p_{r}^{\mu}\frac{\del f^{0}_{r}}{\del x^{\mu}} &- p_{r}^{\mu}p_{r}^{\beta}\Gamma^{\al}_{\mu\bt}\frac{\del f^{0}_{r}+\delta f_{r}}{\del p_{r}^{\al}} + g^{\mu\al}M \del_{\al}M \frac{\del f^{0}_{r}}{\del p_{r}^{\mu}} \nn\\
&+ q_{r} (E^{\al} u^{\bt}-E^{\bt}u^{\al})p_{r\al}\frac{\del f^{0}_{r}}{\del p_{r}^{\bt}}=-\frac{u_{\al}p_{r}^{\al}}{\tau_{c}}\delta f_{r}, \label{A12}
\end{align}
where we have substituted the RTA-collision kernel $C[f_{r}]=-\frac{u_{\al}p_{r}^{\al}}{\tau_{c}}(f_{r}-f^{0}_{r})$ and kept the first order gradients in fluid velocity $u^{\al}$ and the thermodynamic variables $\mu_{r}$ and $T$ in the LHS of Eq.~(\ref{A12}). The $\tau_{c}$ is the average time of collision between quarks. Since our goal is to calculate the electrical conductivity and shear viscosity in the presence of rotation, we only keep the electric-like part $E^{\mu}\equiv F^{\mu\nu}u_{\nu}$ and ignore the magnetic-like part $B^{\mu\nu}$ in the tensor decomposition of $F^{\mu\nu}=E^{\mu} u^{\nu}-E^{\nu}u^{\mu}+B^{\mu\nu}$. Substituting $f^{0}_{r}$ in Eq.~(\ref{A12}) we get,
\begin{widetext}
\bea
-f^{0}(1-f^{0})&\left[\frac{p^{\mu}p^{\al}}{T}\del_{\mu}u_{\al}+p^{\mu}(u^{\al}p_{\al})\del_{\mu}\frac{1}{T}-p^{\mu}\del_{\mu}\frac{\mu}{T}-\Gamma^{\sigma}_{\mu\la}p^{\mu}p^{\la}\frac{u_{\sigma}}{T}-\frac{q}{T}E_{\nu}p^{\nu}+\frac{M}{T}u^{\nu}\del_{\nu}M\right]\nn\\
&-\Gamma^{\sigma}_{\mu\la}p^{\mu}p^{\la}\frac{\del \delta f}{\del p^{\sigma}}=-\frac{u_{\al}p^{\al}}{\tau_{c}}\delta f,\label{A13}
\eea
where we ignored the species label $r$, which will be retained finally during the evaluation of total electric current and shear stress tensor. We rewrite Eq.~(\ref{A13}) in the comoving frame $u^{\mu}\rightarrow (1/\sqrt{g_{00}}~,0)$ by employing the approximations specified in Ref.~\cite{Padhan:2024edf}, 
\bea
-f^{0}(1-f^{0})&\left[-\frac{p^{0}p^{i}}{T}\frac{\del u^{i}}{\del t}-\frac{p^{i}p^{j}}{T}\del_{j}u^{i}+p^{0}p_{0}\frac{\del}{\del t}\frac{1}{T}+p^{i}p_{0}\del_{i}\frac{1}{T}-p^{0}\frac{\del}{\del t}\frac{\mu}{T}-p^{i}\del_{i}\frac{\mu}{T}+\frac{q}{T}\tilde{E}^{i}p^{i}+\frac{M}{T}\frac{\del M}{\del t}\right]\nn\\
&+2 p^{0}\ep^{ijk}p^{j}\Om^{k}\frac{\del \delta f}{\del p^{k}}=-p^{0}\frac{\delta f}{\tau_{c}}~,\label{A14}
\eea
where $i=1 \text{ to } 3$. We will re-express Eq.~(\ref{A14}) in a steady state by retaining the terms that are responsible for electrical conductivity and shear viscosity as follows,
\be
\frac{f^{0}(1-f^{0})}{ET}(p^{i}p^{j}\del_{j}u^{i}-q\tilde{E}^{i}p^{i})+2(\vec{p}\times\vec{\Om})\cdot\frac{\del \delta f}{\del \vec{p}}=-\frac{\delta f}{\tau_{c}},\label{A15}
\ee
where $E=\sqrt{\vec{p}^{2}+M^{2}}$. Solution $\delta f$ of Eq.~(\ref{A15}) can be used to get the expression of current density and shear stress tensor for the system of rotating constituent quarks. The linearity of Eq.~(\ref{A15}) and the anisotropic nature of the Coriolis term containing $\vec{\Om}\equiv\Om \hat{\om}$ (to keep the discussion general we are using a general unit vector $\hat{\om}$ to specify the direction of angular velocity which will be finally identified to be along z-axis, i.e., $\hat{\om}=\hat{k}$), suggest the following ansatz: $\delta f=\sum\limits_{n=0}^{2}A_{n}~X_{n}^{k}~p^{k}+\sum\limits_{n=0}^{4}C_{n}~C_{n}^{kl}~p^{k}p^{l}$, where $C_{n}^{kl}$ and $X_{n}^{k}$ are respectively independent traceless tensor and vector components made up of $U^{kl}$, $\om^{k}\equiv \frac{\Om^{k}}{\Om}$, $\tilde{e}^{k}\equiv \frac{\tilde{E}^{k}}{\tilde{E}}$, $\delta^{kl}$ and $\ep^{ikl}$~\cite{Padhan:2024edf,Dwibedi:2025bdd}. The $A_{n}$ and $C_{n}$ are unknowns that have to be determined by substituting the ansatz in Eq.~(\ref{A15}). Upon substituting $\delta f$ in Eq.~(\ref{A15}) one obtains
\bea
&& A_{0}=\frac{f^{0}(1-f^{0})}{ET}\frac{q\tilde{E}~\tau_{c}}{1+\big(\frac{\tau_c}{\tau_{\Omega}} \big)^2}~,A_{1}=\frac{f^{0}(1-f^{0})}{ET}\frac{q\tilde{E}~\tau_{c}}{1+\big(\frac{\tau_c}{\tau_{\Omega}} \big)^2}\frac{\tau_{c}}{\tau_{\Om}}~,A_{2}=\frac{f^{0}(1-f^{0})}{ET}\frac{q\tilde{E}~\tau_{c}}{1+\big(\frac{\tau_c}{\tau_{\Omega}} \big)^2} \left(\frac{\tau_{c}}{\tau_{\Om}}\right)^{2}(\hat{\omega}\cdot\hat{e})~,\nn\\
&&C_0=-\frac{f^0(1-f^0)}{2ET}\tau_c~,C_1=-\frac{f^0(1-f^0)}{2ET}\frac{\tau_c}{1+4(\tau_c/\tau_\Om)^2}~,C_2=-\frac{f^0(1-f^0)}{2ET}\frac{\tau_c}{1+(\tau_c/\tau_\Om)^2}~,\nn\\
&&C_3=-\frac{f^0(1-f^0)}{ET}\frac{\tau_c(\tau_c/\tau_\Om)}{1+4(\tau_c/\tau_\Om)^2}~,C_4=-\frac{f^0(1-f^0)}{2ET}\frac{\tau_c(\tau_c/\tau_\Om)}{1+4(\tau_c/\tau_\Om)^2}~,\label{A16}
\eea
 where the rotational time period $\tau_{\Om}\equiv 1/(2\Om)$, electric field $\tilde{E}^{k}\equiv\tilde{E}~e^{k}$ and angular velocity $\Om^{k}\equiv\Om~ \om^{k}$. Using the above expressions of $A_{n}$ and $C_{n}$ in $\delta f$ and employing Eq.~(\ref{A7}) and Eq.~(\ref{A9}) we get,
 \bea
 	&&J^{i}=\sum\limits_{r}\frac{g_{r} q_{r}^2}{3T} \int \frac{d^3\Vec{p}}{(2\pi)^3}~\frac{p^2}{E^2}\bigg(\frac{\tau_c}{1+\Big(\frac{\tau_c}{\tau_{\Omega}}\Big)^2}\bigg)\bigg[\delta^{ij}+\Big(\frac{\tau_c}{\tau_{\Omega}}\Big) \epsilon^{ijk}\omega^k + \Big(\frac{\tau_c}{\tau_{\Omega}}\Big)^2\omega^i\omega^j\bigg]\tilde{E}^j f^{0}_{r}(1-f^{0}_{r})~,\label{A17}\\
    &&\pi^{ij} =\sum_{n=0}^{4} C^{ij}_n\sum\limits_{r}\frac{2g_{r}}{15}\int \frac{d^3p}{(2\pi)^3 E}C_n~p^4~,\label{A18}
\eea
\end{widetext}
where we have retained the label $r$ for different constituent quarks. Comparing the above equations of current density and shear stress tensor with the macroscopic version of the expressions provided in Eq.~(\ref{A8}) and Eq.~(\ref{A10}), we obtain the following expressions of conductivities and shear viscosities for the rotating system of quarks,
\bea
 &&\sigma_{n} =\frac{20e^{2}}{9T}\int \frac{d^{3}p}{(2\pi)^3}\frac{\tau_c\big(\frac{\tau_c}{\tau_{\Omega}}\big)^n} {1+\big(\frac{\tau_c}{\tau_{\Omega}}\big)^2}\times \frac{p^2}{E^{2}}f^{0}(1-f^{0}),\nn\\
 &&\text{ with }\sigma^{ij}=\sigma_{0} \delta^{ij} +\sigma_{1}\epsilon^{ijk}\omega^k +\sigma_{2}\omega^i\omega^j~,\label{A19}\\
 &&\eta_0=\frac{8}{5T} \tau_c \int\frac{d^3p}{(2\pi)^3}\frac{p^4}{E^2}f^{0}(1-f^{0})~,\label{A20}\\
&&\eta_1=\frac{8}{5T}\frac{\tau_c}{1+4(\tau_c/\tau_\Om)^2}\int\frac{d^3p}{(2\pi)^3}\frac{p^4}{E^2}f^{0}(1- f^{0})~,\label{A21}\\
&&\eta_2=\frac{8}{5T}\frac{\tau_c}{1+(\tau_c/\tau_\Om)^2}\int\frac{d^3p}{(2\pi)^3}\frac{p^4}{E^2}f^{0}(1-f^{0})~,\label{A22}\\
&&\eta_3=\frac{16}{5T}\frac{\tau_c(\tau_c/\tau_\Om)}{1+4(\tau_c/\tau_\Om)^2}\int\frac{d^3p}{(2\pi)^3}\frac{p^4}{E^2}f^{0}(1-f^{0})~,\label{A23}\\
&&\eta_4=\frac{8}{5T}\frac{\tau_c(\tau_c/\tau_\Om)}{1+(\tau_c/\tau_\Om)^2}\int\frac{d^3p}{(2\pi)^3}\frac{p^4}{E^2}f^{0}(1- f^{0})~,\label{A24}
\eea
where we have multiplied an extra factor of two with the spin and color degeneracy factor $g_{r}$ to incorporate the anti-quarks, which contribute equally to $\mu=0$. We can define the perpendicular $\eta_{\perp}$, parallel $\eta_{||}$ and hall $\eta_{\times}$ viscosity as \cite{Dey:2019axu}, $\eta_{1}\equiv \eta_{\perp}$, $\eta_{2}\equiv \eta_{||}$, and $\eta_{4}\equiv \eta_{\times}$. Similarly, the perpendicular $\sigma_{\perp}$, parallel $\sigma_{||}$ and hall $\sigma_{\times}$ conductivities are defined as $\sigma_{0}\equiv \sigma_{\perp}$, $\sigma_{0}+\sigma_{2}\equiv \sigma_{||}$ and $\sigma_{1}\equiv \sigma_{\times}$. In the absence of rotation, five shear viscosity components merge into one component $\eta_{0}\equiv \eta $, and three electrical conductivity components merge into one component $\sigma_{||}\equiv \sigma$.

Now, we proceed to the next section to describe the results of conductivity and shear viscosity obtained for the rotating system of constituent quarks obtained from the NJL model.
\section{Results and discussion}
\label{sec-result}
\begin{figure}[h!]
\centering
  \includegraphics[width=\linewidth]{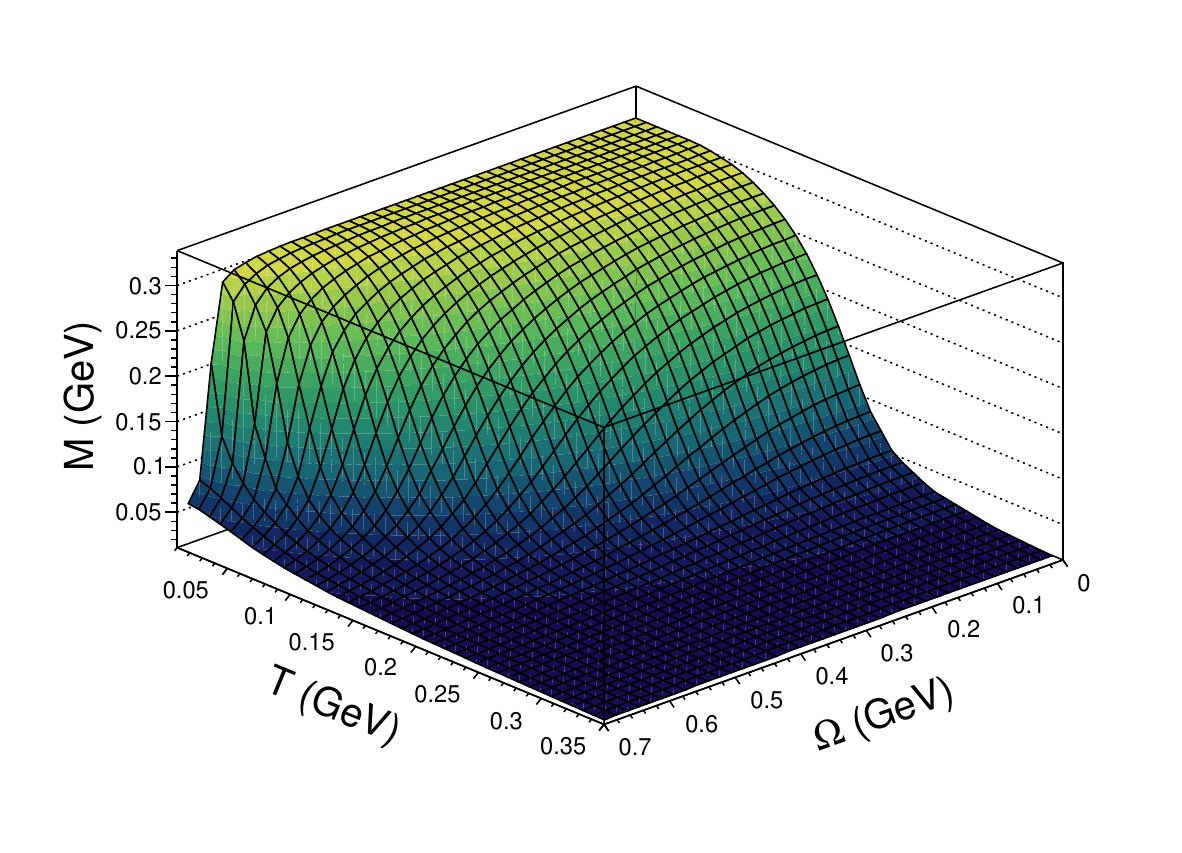}
  \caption{Constituent quark mass ($M$) as a function of angular velocity ($\Omega$) and temperature ($T$).}
\label{fig:Mass}
\end{figure}
For the numerical estimation of constituent quark mass and thermodynamic equation of state (EoS) from the NJL model, we used the following parameter set. To fit pion mass and decay constant in the vacuum, we consider current quark mass $m_u = m_d = 5.5$~MeV, cut off to regularize the vacuum term is $\Lambda = 651$~MeV, and the scalar coupling $G_s = 5.04 \times 10^{-6}$~MeV$^{-2}$~\cite{Klevansky:1992qe}. Moreover, all the results are at position $\rho= 0.1$~GeV$^{-1}$~\cite{Jiang:2016wvv}, and for vanishing baryon chemical potential, $\mu_B = 0$. Angular velocity $\Omega$ is a free parameter bounded by the causality relation $\rho \Omega < 1$. 

In Fig.~(\ref{fig:Mass}), we depict constituent quark mass ($M$) as a function of angular velocity ($\Omega$) and temperature ($T$). As temperature $T$ increases, the value of the constituent quark mass reduces from $M=313$~MeV to $M=m_{u}=m_{d}=5.5$ MeV. This is the standard representation of a chiral phase transition from the chiral symmetry-breaking phase to the restored phase. The temperature where the change of mass ($\frac{dM}{dT}$) becomes maximum is commonly known as the chiral phase transition temperature $T_{c}$. Interestingly, we notice that $M$ also reduces in the rotating frame with increasing $\Omega$. As a result, rotation can reduce the chiral transition temperature ($T_c$). However, in HIC, the vorticity or $\Omega$ is substantially low ($\approx 0.01$~GeV~\cite{XuGuangHuang:2020dtn,PhysRevC.95.054915,XuGuangHuang2016,Jiang2016}) to make any significant impact in $T_c$.  Even if we increase $\rho$, as shown in Ref.~\cite{Jiang:2016wvv}, at such low $\Omega$, effect in $M$ or $T_c$ is negligible. Though phenomenological values of $\Omega$ are not high enough to have an impact on $T_{c}$, different thermodynamical and transport quantities can be modified due to the modification of $M(\Omega)$. We will see these quantities next.

%%%%%%%%%%%%%%%%%%%%%%%%%%
\begin{figure}[h!]
\centering
  \includegraphics[width=\linewidth]{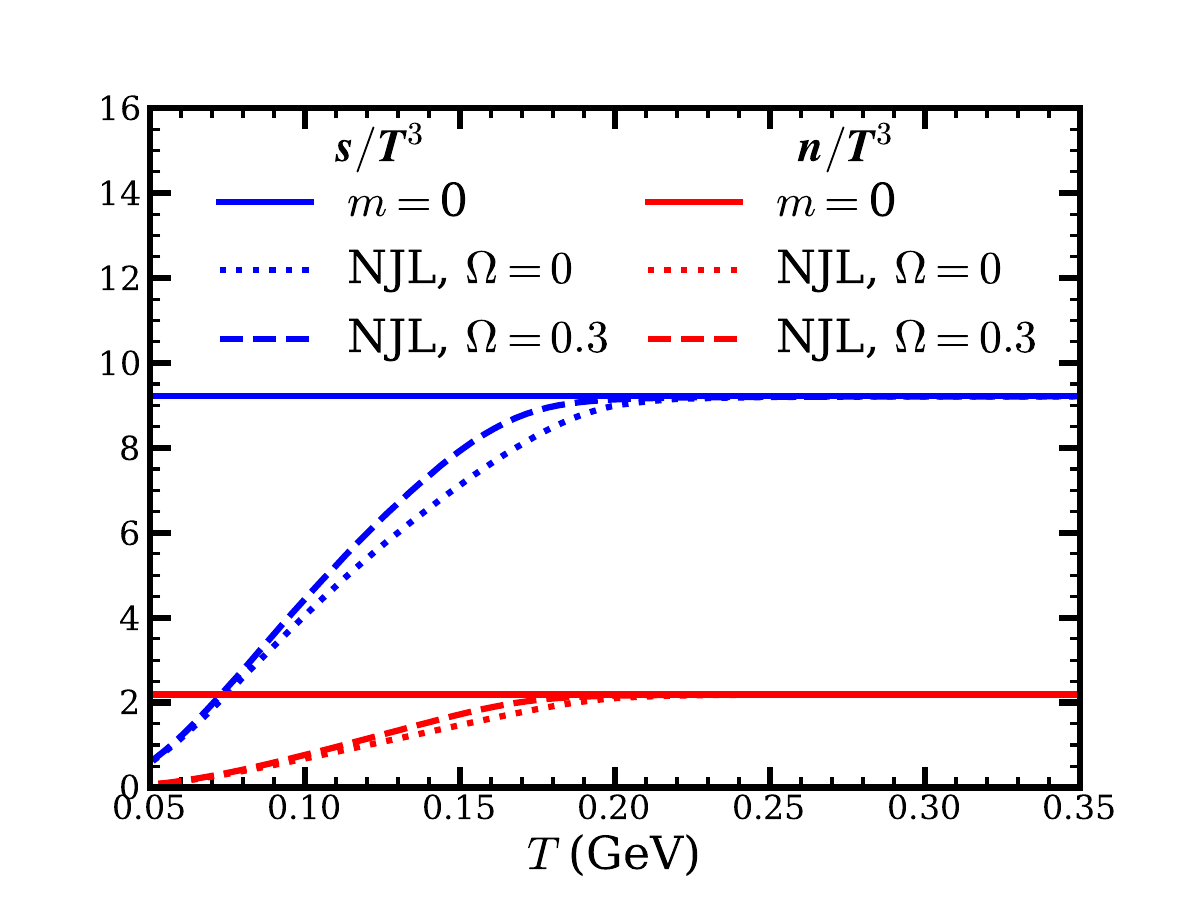}
  \caption{Variation of thermodynamic variables: entropy $s$ (blue) and total number density $n$ (red) as a function of temperature $T$ at different values of  $\Omega$.}
\label{fig:Thermo}
\end{figure}
In Fig.~(\ref{fig:Thermo}), we plot the entropy density ($s$) and total quark number density ($n$) scaled by $T$ as a function of temperature at two different values of $\Omega = 0, 0.3$ GeV.  The dotted and dashed lines represent NJL model estimations obtained from the grand potential given in Eq.~(\ref{A3}) and with the help of the Euler thermodynamic relation. Solid lines represent results for a massless, non-interacting gas of quarks. We can see that at a higher angular velocity, the thermodynamic quantities are enhanced (dashed lines lie above the dotted ones). This is a result of a decrease in quark condensate with $\Omega$. Besides thermodynamic EoS, the behavior of $s$ is important to analyze the fluidity measure, which is a ratio of shear viscosity to entropy density ($\eta/s$). On the other hand, the total quark number density $n$ is an important quantity for the determination of the average relaxation time $\tau_{c}$ of the particles.
%We will need the entropy to see the numerical magnitude of $\eta/s$ in the NJL framework with $\tau_{c}$ calibrated from previous field theoretical studies. Presently, we will focus either on the departure of the viscosities and conductivities from their usual components, i.e., $\eta_{0}\equiv \eta$ and $\sigma_{||}\equiv\sigma$ at a constant relaxation time or on the thermodynamic phase part of the viscosities and conductivities by normalizing them with relaxation time and temperature.

%%%%%%%%%%%%%%%%%%%%%%%%
\begin{figure*}
%\end{figure*}[h!]
    \centering
    \begin{subfigure}[b]{0.45\textwidth}
         \centering
         \includegraphics[width=\linewidth]{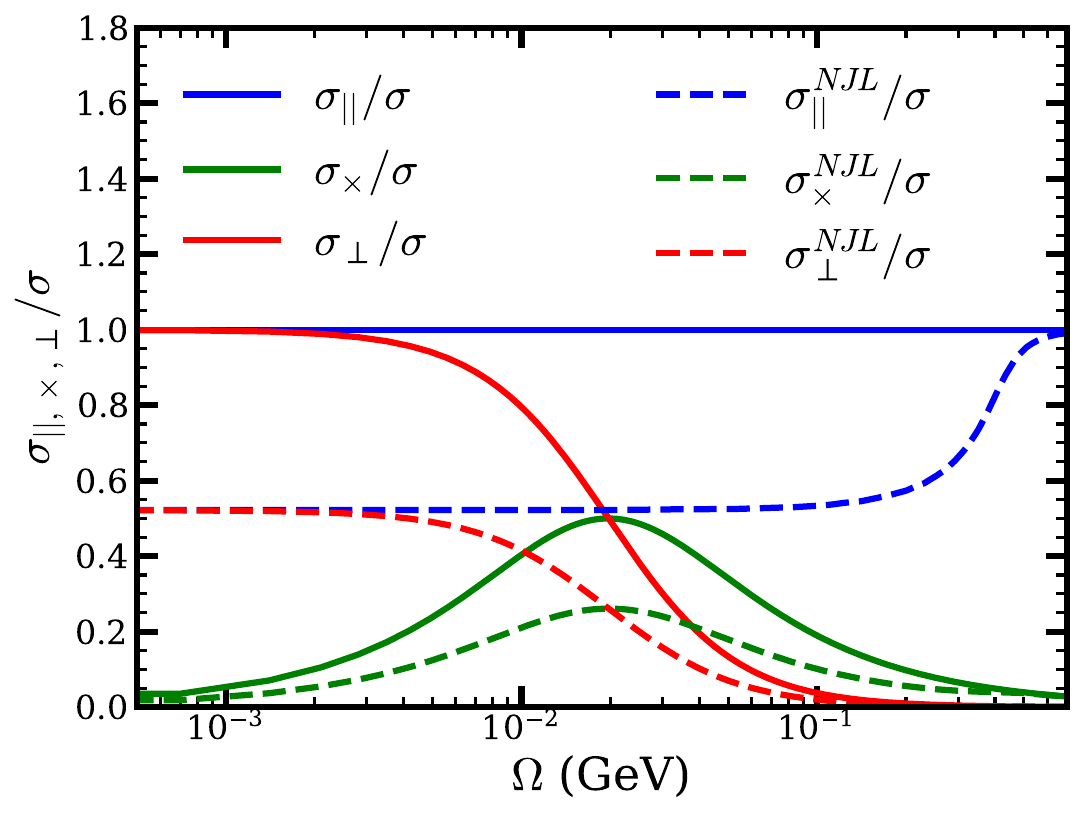}
        \caption{Electrical conductivity components against $\Omega$ at $T=150$ MeV and $\tau_{c}=5$ fm}
    \label{nsigmaOm}
    \end{subfigure}
    \hfill
    \begin{subfigure}[b]{0.45\textwidth}
         \centering
        \includegraphics[width=\linewidth]{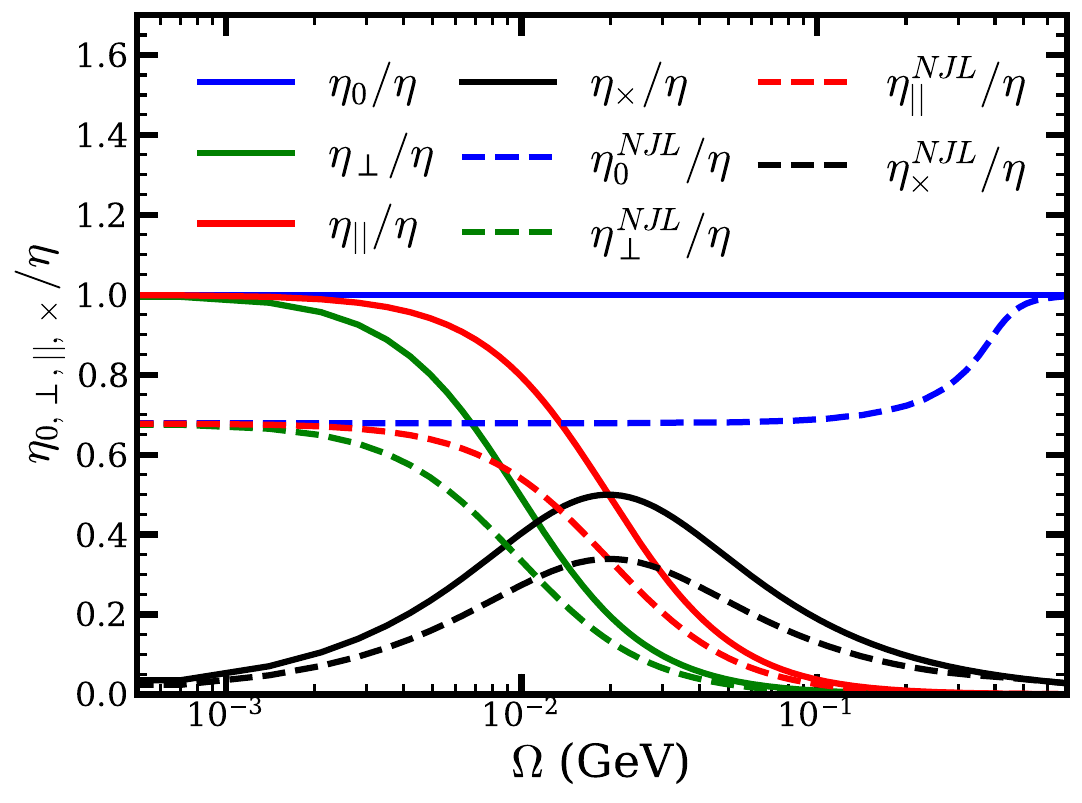}
        \caption{Shear viscosity components against $\Omega$ at $T=150$ MeV and $\tau_{c}=5$ fm}
    \label{nsigmaT}
    \end{subfigure}
    \caption{Anisotropic transport coefficient components normalized to the corresponding isotropic coefficient, plotted as a function of angular velocity $\Omega$ at $T=150$ MeV. $\sigma_{||}\equiv \sigma $ and $\eta_0\equiv \eta$ are the isotropic conductivity and viscosity, respectively, in the non-rotating frame.}
    \label{fig:Nconductivities}
\end{figure*}
Fig.~(\ref{nsigmaOm}) shows the components of electrical conductivity $\sigma_{||,\perp,\times}$ in the rotating frame, normalized by the isotropic conductivity in the absence of $\Omega$ (here $\sigma= \sigma_{||}$). We plot their variation with angular velocity for a fixed temperature of $T = 150$~MeV. The parallel ($\parallel$), perpendicular ($\perp$), and Hall ($\times$) components are represented by the blue, red, and green lines, respectively. The dashed and solid lines are NJL and massless non-interacting estimations, respectively. $\sigma$ is taken for massless gas. Here, we consider a constant relaxation time $\tau_c=5$ fm. By decreasing $\Omega$, the Hall-type component vanishes, and the other two components $\perp$ and $\parallel$ merge as a single component, which means the anisotropy of the conductivity tensor disappears (as expected). At low $\Omega$, massless results differ from the NJL estimation, which carries a non-zero constituent mass due to non-zero quark condensate. However, at very high $\Omega$, massless results recover due to chiral restoration. The Hall-type component in the rotating frame has a significant contribution at zero chemical potential, unlike in the presence of a magnetic field. The Lorentz force induced by a magnetic field generates anisotropic transport, with positively and negatively charged particles contributing oppositely to the Hall component. In contrast, the Coriolis force in a rotating frame does not distinguish particles based on their charge. As a result, Hall-type components can play a significant role in transport phenomena in rotating systems. The peak in the Hall component appears at a $\Omega$ value where rotational relaxation time ($\tau_\Omega$) equals thermal relaxation time, which here is $\tau_c =5$ fm (cf. Eq.~(\ref{A24})), beacause of the factor $\frac{\tau_c/\tau_\Omega}{1 + (\tau_c/\tau_\Omega)^2}$ in the expression of $\sigma_1$.   
Fig.~(\ref{nsigmaT}) represents the same as Fig.~(\ref{nsigmaOm}) but for shear viscosity components. There are few differences in the expressions of $\sigma_{||,\perp,\times}$ and $\eta_{||,\perp,\times}$, which can be understood from the earlier formalism section, e.g., $\sigma_{||}=\sigma$ but $\eta_{||}\neq \eta$, rather $\eta_{0}=\eta$. That is why we have drawn viscosity components by normalizing them with $\eta_{0}$ or $\eta$. 

%%%%%%%%%%%%%%%%%%%%%%%%%%%%%%%%
\begin{figure*}
    \centering
    \begin{subfigure}{.45\textwidth}
        \centering
        \includegraphics[width=\linewidth]{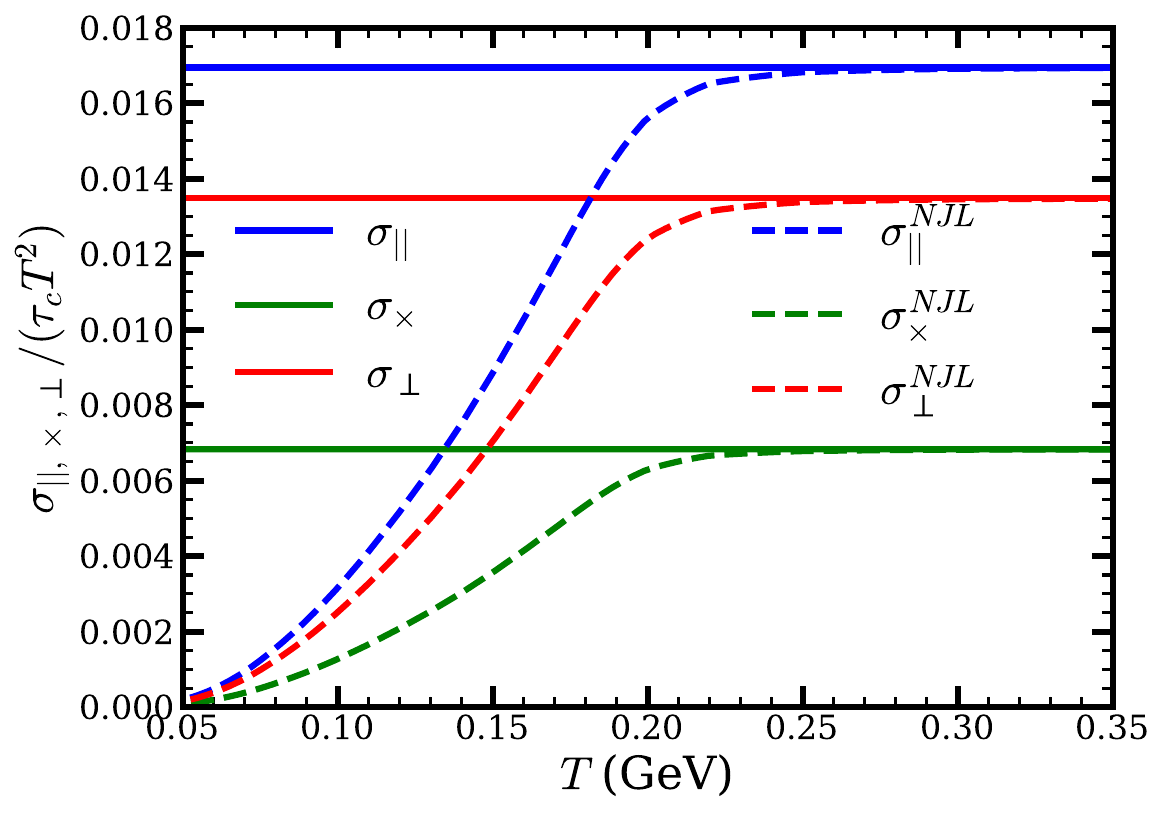}
        \caption{Electrical conductivity components vs $T$ at $\Omega=0.01$ GeV and $\tau_{c}=5$ fm}
        \label{netaOm}
    \end{subfigure}
    \hfill
    \begin{subfigure}{.45\textwidth}
        \centering
        \includegraphics[width=\linewidth]{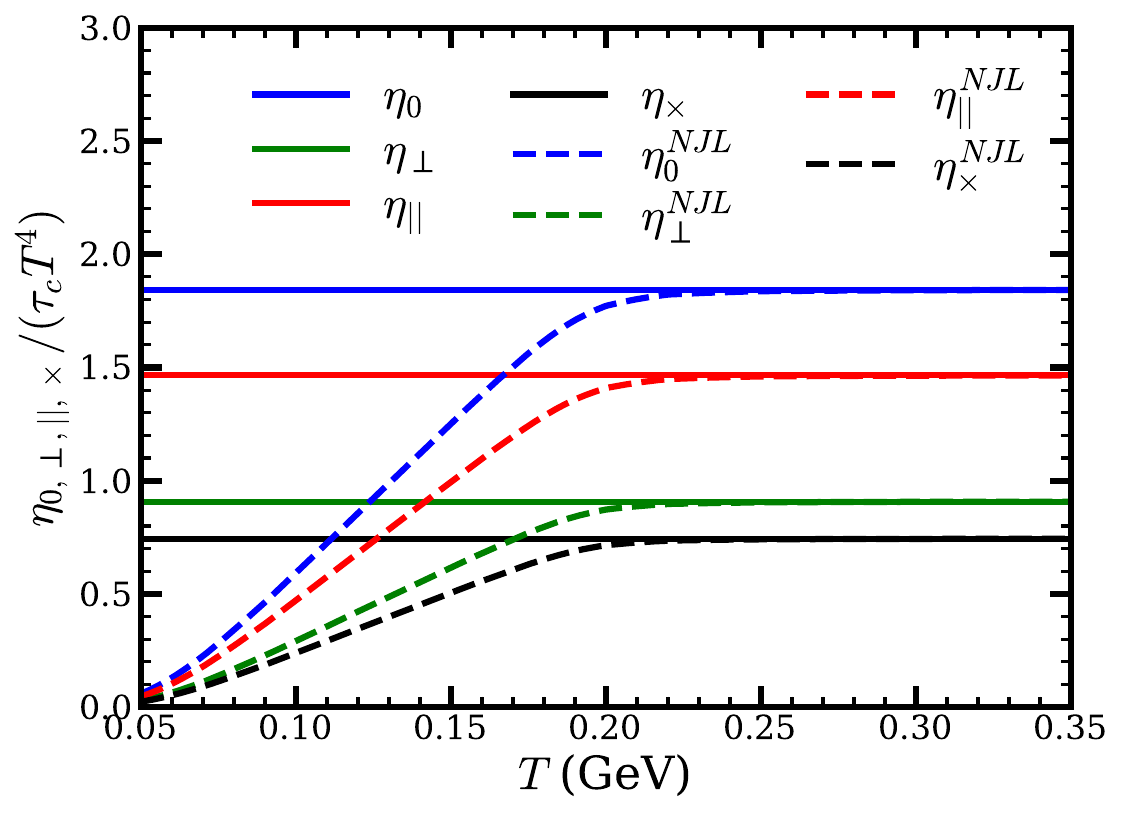}
        \caption{Shear viscosity components vs $T$ at $\Omega=0.01$ GeV and $\tau_{c}=5$ fm}
        \label{netaT}
    \end{subfigure}
    \caption{Components of transport coefficients scaled by relaxation time ($\tau_c$) and temperature ($T$) are plotted against $T$ at $\Omega=0.01$ GeV}
    \label{fig:Nviscosities}
\end{figure*}
Fig.~(\ref{netaOm}) depicts the variation of conductivity components with $T$ at $\Omega = 0.01$~GeV. Conductivity is made dimensionless by dividing with $\tau_c T^2$. To highlight the effect of rotating thermodynamics, we set a fixed value of the relaxation time $\tau_c =5$~fm. The conductivity increases with $T$ as the constituent quark mass decreases. At high $T$, chirality restores, and the NJL estimations converge to the massless case. A similar trend is observed for the shear viscosity components, as shown in Fig.~(\ref{netaT}). We consider $\Omega=0.01$ GeV, which is within the phenomenological values. Here we notice a substantial difference between parallel and perpendicular components (reflecting anisotropy) and a non-zero Hall component. 
%Moreover, in the NJL estimation, the anisotropy in transport coefficients which is seperation of individual components, increases as the temperature increase. However, for massless estimation anisotropy remains independent of the temperature. It is only appearent. To realize actual scenario we have to normalize it with absolute value of the components. 

Now, for a realistic estimation of the transport coefficients of the rotating quark matter produced in HIC, we will tune our relaxation time $\tau_{c}$ to fit earlier theoretical estimations~\cite{Cassing:2013iz,Fernandez-Fraile:2009eug,Greif2014,Marty:2013ita,Puglisi:2014pda,Amato2013,Gorenstein:2007mw,Itakura:2007mx,Plumari:2012ep, Noronha-Hostler:2008kkf} and eventually we will get a numerical band of $\tau_{c}$ from those earlier works~\cite{Cassing:2013iz,Fernandez-Fraile:2009eug,Greif2014,Marty:2013ita,Puglisi:2014pda,Amato2013,Gorenstein:2007mw,Itakura:2007mx,Plumari:2012ep, Noronha-Hostler:2008kkf}. For the temperature $T> 170$ MeV, where one expects deconfined quark matter, we assume $\tau_{c}$ to be a tunable constant. For the temperature $T<170$ MeV, we calculate $\tau_{c}$ using a hard-sphere scattering relation,  
\be
\tau_{c}=\frac{1}{n~\pi a^{2}~v_{\text{av}}}~,\label{tauh}
\ee
where $a$ is the scattering length treated here as a parameter. The $n$ and $v_{\text{av}}$ are, respectively, the total quark density and thermal average velocity of the quarks given by,
\bea
    && n = 24\int{\frac{d^{3}\vec{p}}{{(2\pi})^{3}}} ~f^{0},\nn\\
    && v_{\text{av}} = \frac{\int{\frac{d^{3}\vec{p}}{{(2\pi})^{3}}} \frac{p}{E}~f^0}{\int{\frac{d^{3}\vec{p}}{{(2\pi})^{3}}} ~f^{0}},
\eea
where $f^{0}=1/[e^{E/T}+1]$ is the Fermi-Dirac distribution for the constituent quarks with energy $E=\sqrt{\vec{p}^{2}+M^{2}}$. The realistic range for the $\tau_{c}$ in the whole temperature range is found by covering the numerical magnitudes of the $\sigma/T$ and $\eta/s$ obtained by different authors~\cite{Cassing:2013iz,Fernandez-Fraile:2009eug,Greif2014,Marty:2013ita,Puglisi:2014pda,Amato2013,Gorenstein:2007mw,Itakura:2007mx,Plumari:2012ep, Noronha-Hostler:2008kkf} at $\Om=0$ with the $\sigma$, $\eta$ and $s$ given by,
\bea
    &&\sigma = \frac{20 e^{2}}{9T}\tau_{c}\int \frac{d^{3}p}{(2\pi)^{3}} ~\frac{p^{2}}{E^{2}}~ f^{0}(1-f^{0})\nn\\
    &&\eta = \frac{8}{5T}\tau_{c}\int \frac{d^{3}p}{(2\pi)^{3}} ~\frac{p^{4}}{E^{2}} ~f^{0}(1-f^{0})\nn
\eea
and entropy $s(\Om\rightarrow 0)$ obtained from the grand potential in Eq.~(\ref{A3}).
%%%%%%%%%%%%%%%%%%%%%%%%%%%
\begin{figure*}
    \centering
    \begin{subfigure}{.45\textwidth}
        \centering
        \includegraphics[width=\linewidth]{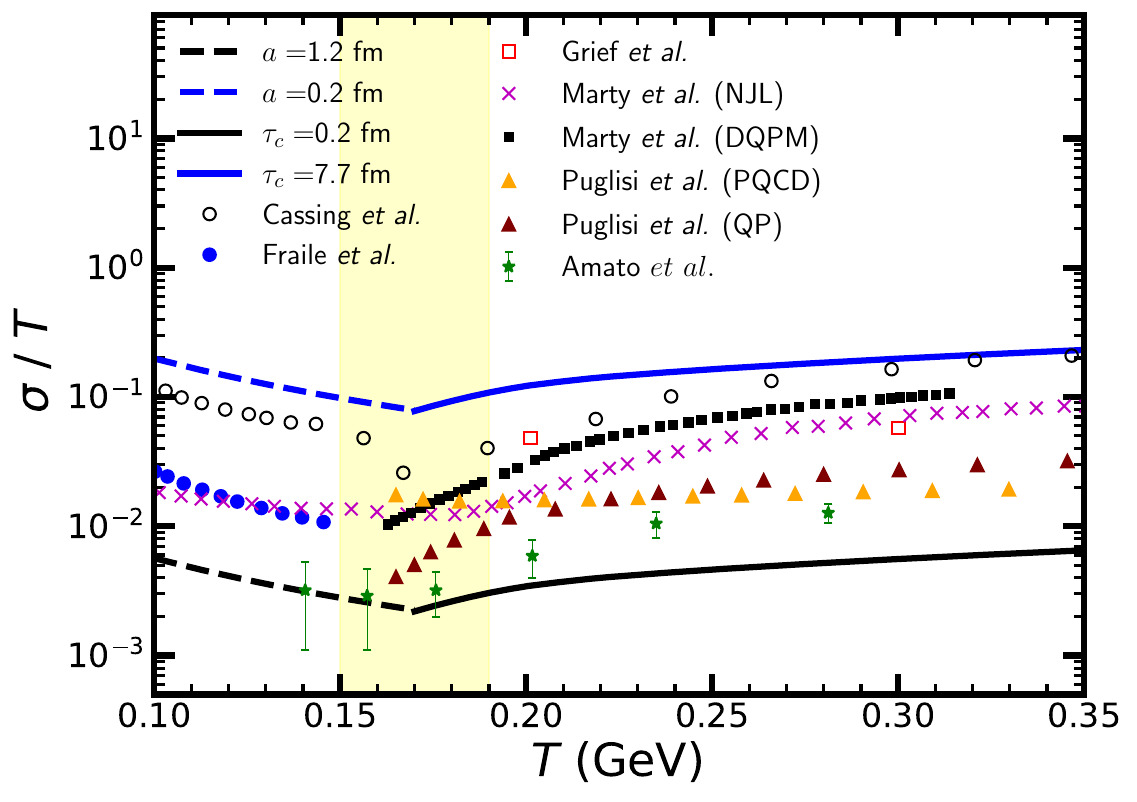}
        \caption{Calibration of $\tau_{c}$ from comparison with existing calculation of conductivity at $\Omega=0$. We used the results of the following papers, Cassing \textit{et al.} (PHSD)~\cite{Cassing:2013iz}, Fraile \textit{et al.} (ChPT)~\cite{Fernandez-Fraile:2009eug},Greif \textit{et al.} (BAMPS)~\cite{Greif2014}, Marty \textit{et al.} (NJL)~\cite{Marty:2013ita}, Marty \textit{et al.} (DQPM)~\cite{Marty:2013ita}, Puglisi \textit{et al.} (PQCD)~\cite{Puglisi:2014pda}, Puglisi \textit{et al.} (QP)~\cite{Puglisi:2014pda}, Amato \textit{et al.} (LQCD)~\cite{Amato2013}}
        \label{sigmatau}
    \end{subfigure}
    \hfill
    \begin{subfigure}{0.45\textwidth}
        \centering
        \includegraphics[width=\linewidth]{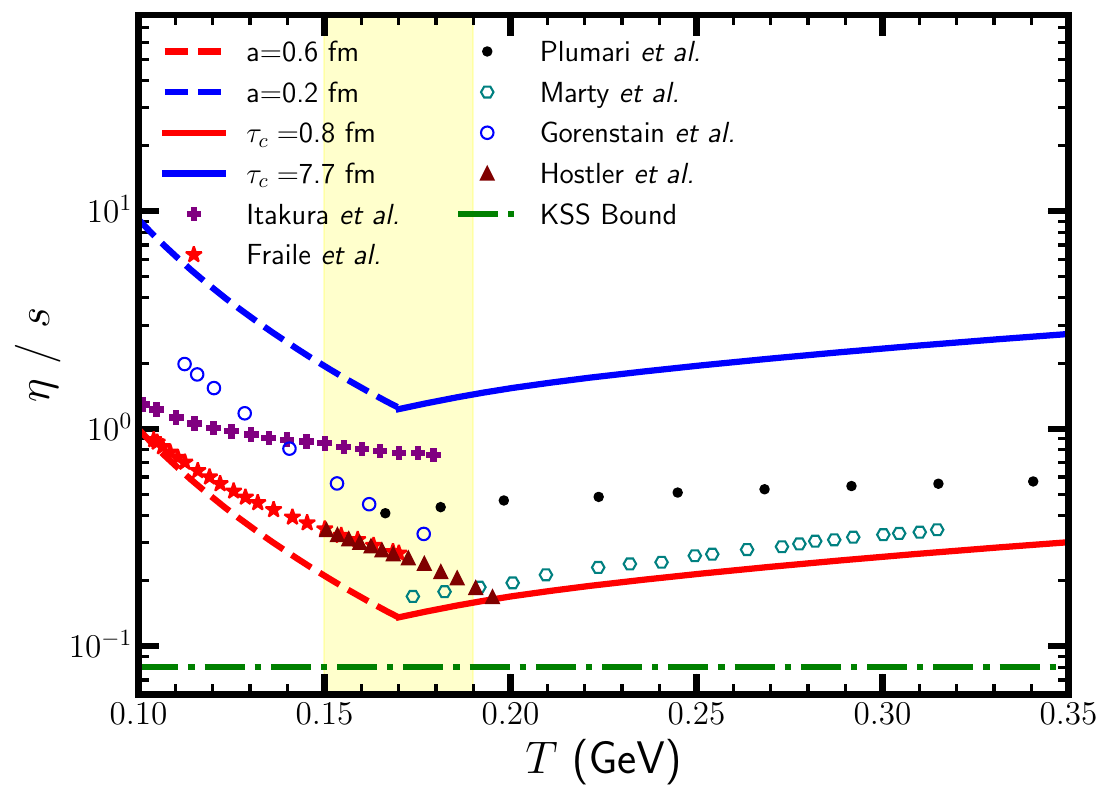}
        \caption{Calibration of $\tau_{c}$ from comparison with existing calculation of viscosity at $\Omega=0$. We used the results of the following papers:  Itakura \textit{et al.} (pion gas)~\cite{Itakura:2007mx}, Fraile \textit{et al.} (ChPT) ~\cite{Fernandez-Fraile:2009eug}, Plumari \textit{et al.} ~\cite{Plumari:2012ep}, Marty \textit{et al.} (DQPM)~\cite{Marty:2013ita},Gorenstein \textit{et al.} (VDWHRG)~\cite{Gorenstein:2007mw}, Hostler \textit{et al.}~\cite{Noronha-Hostler:2008kkf}.}
        \label{etatau}
    \end{subfigure}
    \caption{Scaled conductivity and viscosity as a function of temperature from various model calculations and first principle LQCD result, to obtain a phenomenologically relevant $\tau_c$.}
    \label{fig:caltauc}
\end{figure*}
%%%%%%%%%%%%%%%%%%%%%%%%%%%%%%%%%
In Fig.~(\ref{sigmatau}), we have calibrated the scattering length $a$ for the temperature domain $T<170$ MeV to cover the numerical band of the $\sigma/T$ obtained in the Refs.~\cite {Cassing:2013iz,Fernandez-Fraile:2009eug,Marty:2013ita}. Similarly, for the temperature range $T>170$ MeV, we tuned the relaxation time $\tau_{c}$ to cover the numerical values of $\sigma$. It is evident from Fig.~(\ref{sigmatau}) that for $\tau_{c}\in (0.2,7.7)$ fm and $a\in(0.2,1.2)$ fm, the NJL model estimations cover the results of many authors in the whole temperature range. The same calibration procedure for $\eta/s$ in Fig.~(\ref{etatau}) results in $\tau_{c}\in(0.8,7.7)$ fm and $a\in (0.2, 0.6)$ fm. The shaded yellow band in Fig.~(\ref{sigmatau}) and 
Fig.~(\ref{etatau}) around $T=170$ MeV is to remind the readers that our main focus in the present paper is to see the overall trend of shear viscosity and electrical conductivity in the high-temperature (chirally-symmetry restored) and low-temperature (chirally-symmetry broken) regime. Although we have smoothly matched the curves of two phases at T = 170 MeV, the analysis mainly focuses on the order of magnitude and qualitative behavior of shear viscosity and electrical conductivity components outside this shaded region. 
%%%%%%%%%%%%%%%%%%%%%%%%%%%%%%%%%%
    \begin{figure}
        \centering
        \includegraphics[width=\linewidth]{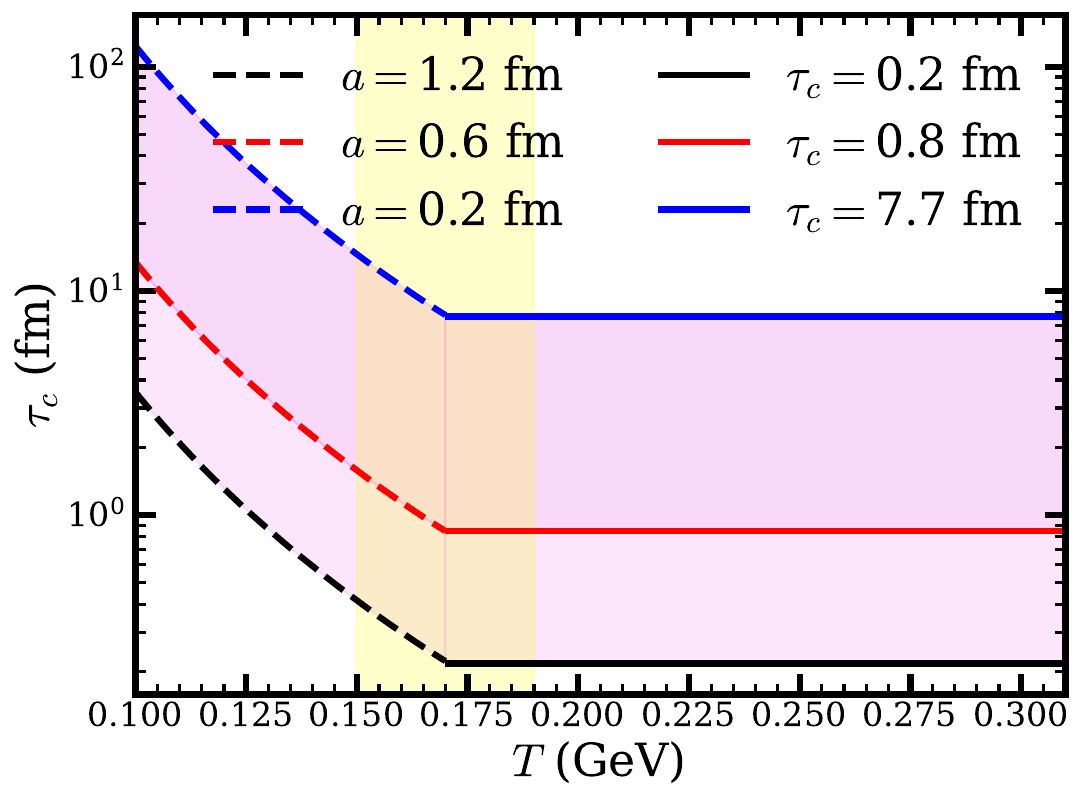}
        \caption{Band of $\tau_{c}$ obtained from calibration of $\sigma$ and $\eta$ in the absence of rotation.}
        \label{fig:tauc}
    \end{figure}
    
In Fig.~(\ref{fig:tauc}), we have shown the band of $\tau_c$ covering the existing model calculations of $\sigma$ and $\eta$ in the absence of rotation. Below $T = 170$~MeV, the scattering length $a = 0.2 - 1.2$~fm cover all the results for $\sigma/T$, and $a = 0.2 - 0.6$~fm for $\eta/s$. Above $T = 170$~MeV, $\tau_c$ vary from $\sim 0.2 - 7.7$~fm for $\sigma/T$, from $0.8 - 7.7$~fm for $\eta/s$.   

%%%%%%%%%%%%%%%%%%%%%%%%%%%%
\begin{figure*}
    \centering
    \begin{subfigure}{.45\textwidth}
        \centering
        \includegraphics[width=\linewidth]{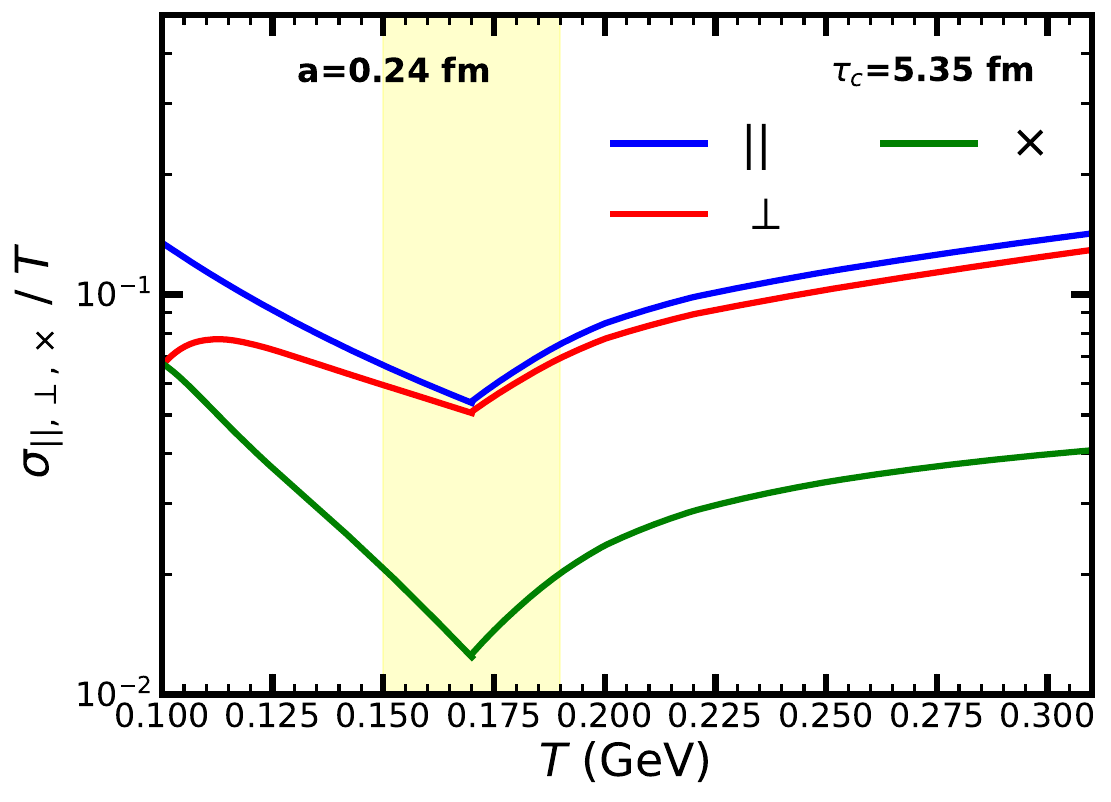}
        \caption{(Color online) The variation of anisotropic components of conductivity ($\sigma_{\perp,\times}/T$) against
temperature by taking a temperature-dependent $\Omega(T)$ is compared with isotropic conductivity $\sigma_{||}/T$ in the absence of rotation.}
        \label{sigmaT}
    \end{subfigure}
    \hfill
    \begin{subfigure}{.45\textwidth}
        \centering
        \includegraphics[width=\linewidth]{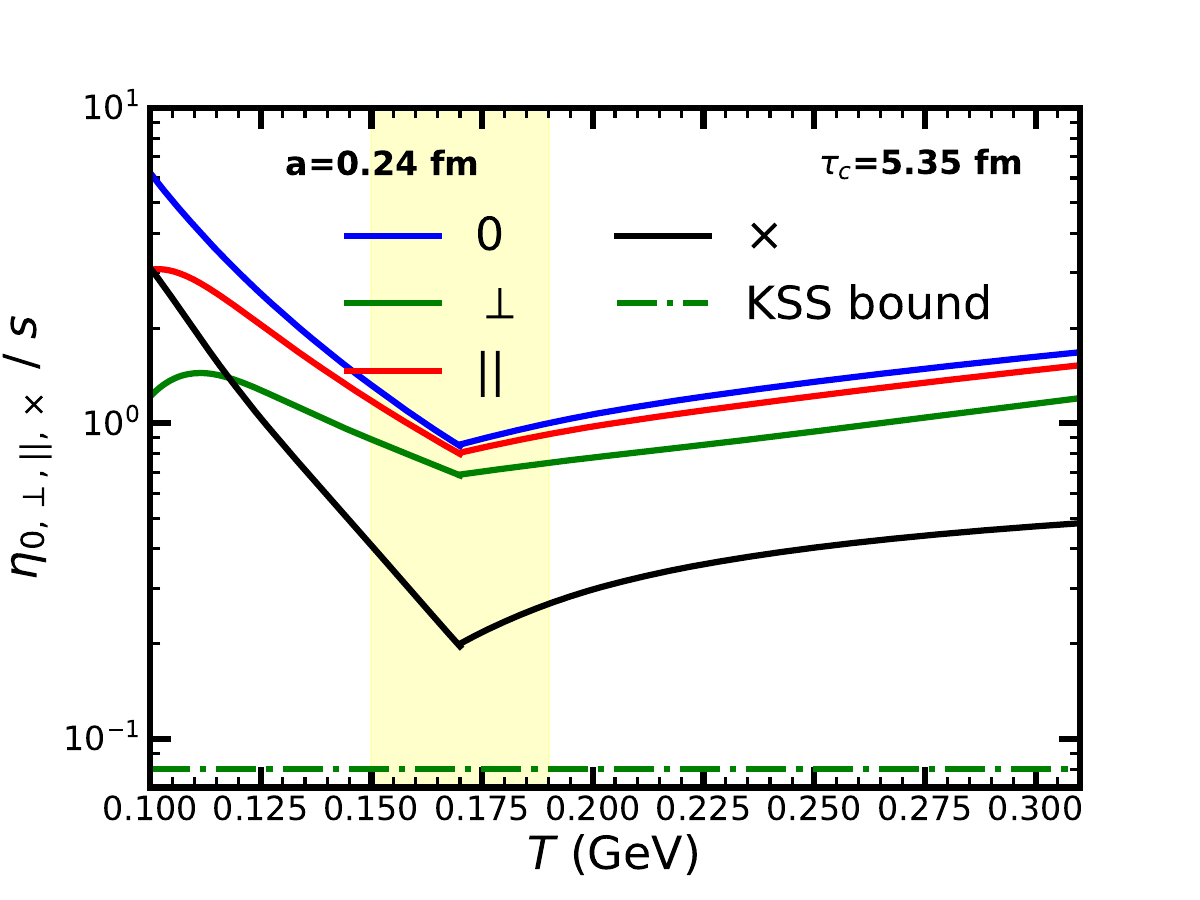}
        \caption{(Color online) The variation of anisotropic shear viscosity to entropy density ratio ($\eta_{\perp,||,\times}/T$) against
temperature by taking a temperature-dependent $\Omega(T)$ is compared with isotropic component $\eta_{0}/T$ in the absence of rotation.}
        \label{etaT}
    \end{subfigure}
    \caption{Estimation of $\sigma/T$ and $\eta/s$ using phenomenological $\Omega(T)$ for matter created in HIC.}
    \label{fig:sigma_eta}
\end{figure*}
After calibrating the $\tau_{c}$ in the whole temperature range, we give a realistic estimate of the anisotropic conductivities and viscosities as a function of $T$. The anisotropic shear viscosity and conductivity components depend on $T$ as well as $\Omega$. One can obtain a temperature-dependent angular velocity $\Om(T)$ faced by the fireball during its evolution as follows. The produced fireball in HIC expands with time, thereby increasing its volume. Assuming an isoentropic expansion and angular momentum conservation during each snapshot of its evolution, one can expect a decreasing temperature and angular velocity profile with fireball evolution time. To express the above argument quantitatively we take the time-dependent angular velocity profile given by Jiang \textit{et al.} \cite{Jiang2016} using multiphase transport (AMPT) model simulations~\footnote{In Refs.~\cite{Jiang2016}, $y-$ axis is chosen perpendicular to the reaction plane whereas we have defined $z-$ axis to be perpendicular to the reaction plane.}, 
\begin{eqnarray}
   \Omega (t, b, \sqrt{s_{NN}}) = \frac{1}{2}[A(b, \sqrt{s_{NN}})  
	+ B(b, \sqrt{s_{NN}})\nonumber\\
   \times (0.58t)^{0.35} e^{-0.58t}],\label{vort}  
\end{eqnarray} 
with the two coefficients \( A \) and \( B \) given by  
\begin{align}
	A &= \left[ e^{-0.016 b \sqrt{s_{NN}}} + 1 \right] \times \tanh(0.28 \, b)\nonumber\\
    &\times \left[ 0.001775 \tanh(3 - 0.015 \sqrt{s_{NN}}) + 0.0128 \right], \nonumber\\
	B &= \left[ e^{-0.016 b \sqrt{s_{NN}}} + 1 \right] \times \left[ 0.02388 \, b + 0.01203 \right]\nonumber\\
    &\times \left[ 1.751 - \tanh(0.01 \sqrt{s_{NN}}) \right]\nonumber,
\end{align}
where beam energy \( \sqrt{s_{NN}} \) is measured in GeV, impact parameter \( b \) in fm, \( t \) in fm/c, and \( \Omega \) in fm\(^{-1}\). Finally, to get the temperature-dependent $\Omega(T)$, a crude temperature-time link can be made with the help of Bjorken’s scaling solution $T(t) = T_0 \left( \frac{t}{t_0} \right)^{-1/3}$. In this analysis, we consider a fireball evolution corresponding to a collision at beam energy \(\sqrt{S_{NN}}=\) 200 GeV with an impact parameter \(b=\) 5 fm. The initial conditions are taken as an initial temperature \(T_0=0.31\) GeV and initial time \(t_0=\) 0.32 fm/c. 

In Fig.~(\ref{sigmaT}) and Fig.~(\ref{etaT}), we respectively show the variation of the scaled conductivities $\sigma_{\perp,\times}/T$ and shear viscosity to entropy density ratio $\eta_{\perp,||,\times}/s$ against temperature. We choose an intermediate value of $\tau_{c}=5.35$ fm for quark temperature domain and scattering length $a=0.24$ fm for hadron temperature domain, which falls in the band obtained in Fig.~(\ref{fig:tauc}). The temperature-dependent $\Omega(T)$ is used in the expressions~(\ref{A19}) to~(\ref{A24}) for the computation of the anisotropic conductivity and viscosity. The isotropic component of conductivity ($\sigma_{||}$) and viscosity ($\eta_{0}$) have been shown for the purpose of comparison with other anisotropic components. In Fig.~(\ref{etaT}) we also display the conjectured KSS bound~\cite{Kovtun:2004de}, which is considered as a lower bound of the $\eta/s$. Since our aim is to see the modification of $\eta_{||,\perp,\times}/s$ and $\sigma_{\perp,\times}/T$ due to phenomenological $\Omega(T)$ of expanding quark and hadronic matter, we have first tuned the valley-shaped temperature profile of $\sigma/T$ and $\eta/s$ with an approximate numerical band, and after it we are proceeding to notice modifications in $\eta_{||,\perp,\times}/s$ and $\sigma_{\perp,\times}/T$. We observe in Fig.~(\ref{sigmaT}) and Fig.~(\ref{etaT}) that the perpendicular components of normalized viscosity and conductivity become smaller than their parallel component, which reflects the building of anisotropic transportation. On the other hand, the non-zero Hall components of transport coefficients are observed. Interestingly, all components, follow the qualitative valley-shaped temperature profile due to consideration of phenomenological $\Omega(T)$. In this regard, we have also noticed that these shapes can be spoiled by using a non-phenomenological constant $\Omega$ in the entire temperature range, which we do not show here. We can see a lesser anisotropy ($\sigma_{||}-\sigma_{\perp}$ or $\eta_{0}-\eta_{\perp,||}$) at higher temperatures ($T>170$ MeV) whose amount increases with decreasing temperature ($T<170$ MeV) and remain significant as the system approaches kinetic freeze-out ($T\approx 100$ MeV).

In a nutshell, we observe an appreciable magnitude of anisotropic conductivities and viscosities for a realistic range of average vorticity expected in HIC~\cite{XuGuangHuang:2020dtn, PhysRevC.95.054915,XuGuangHuang2016,Jiang2016}. These transport coefficients can significantly affect the anisotropic flows and/or particle spectra. Moreover, photon or dilepton spectra are linked to the conductivity of the nuclear medium~\cite{Fernandez-Fraile:2009eug}, so one may also expect the anisotropy in the dilepton spectra because of the anisotropic conductivities.  Our immediate future plan is to study in detail the above-mentioned phenomenological aspects of the anisotropy produced due to rotation.
\section{Summary}
\label{Section-Summary}
In peripheral heavy-ion collisions, a fraction of initial orbital angular momentum can be transferred to the locally thermalized QCD medium in local vorticity. To explain observables such as global and local or longitudinal polarization, an understanding of thermodynamic EoS and transport properties of rotating QCD matter is essential. In this work we studied transport properties such as electrical conductivity and shear viscosity, considering rotating QCD matter. The thermodynamic EoS is obtained using a two-flavor NJL model in a rotating frame. For thermodynamical consistency, we considered a close to equilibrium solution, and all results are obtained close to the rotation axis ($\rho = 0.1~\rm GeV^{-1}$). Moreover, for angular velocity expected in HIC ($\Omega \approx 0.01$~GeV), the temperature in the rotating frame does not deviate much within the causality bound. 

Quark condensate in the NJL model with a fixed coupling constant value ($G_{S}$) reduces the pseudo-critical temperature with rotation. As a result, the constituent quark mass also decreases with rotation, and chiral restoration can be achieved at very high $\Omega$ even at low $T$. Using the kinetic theory framework, we study the transport properties of rotating quark matter with the quasiparticle description from the NJL model. Transport coefficients are obtained by solving the Boltzmann transport equation in a rotating frame, where the Coriolis force can play an important role in building anisotropic transport coefficients. We obtained five independent shear viscosity components and three conductivity components. Similar results were also obtained for transport coefficients at a finite magnetic field due to the Lorentz force. The effect of rotation on the transport coefficients comes from two sources. One is splitting the isotropic transport coefficients into anisotropic multi-components. It happens because the rotational relaxation time $\tau_\Omega$ along with thermal relaxation  $\tau_{c}$, modifies the effective relaxation time in different directions. 
The second is via EoS, which is modified due to improved chiral condensate. This effect increases the value of transport coefficients -  conductivity and viscosity as the constituent quark mass decreases with $\Omega$. Our investigation shows that the effect of the second source of $\Omega$ in the calculation of thermodynamic and transport coefficients can be safely ignored in the phenomenological range of $\Omega$.

Due to the lack of direct experimental measurement of transport coefficients, these are loosely restricted quantities. For phenomenological significance, we modeled a band of the relaxation time ($\tau_c$) to match existing results of the isotropic transport coefficients in the absence of rotation. Below the pseudo-critical temperature, $\tau_c$ is calculated using a hard sphere scattering formalism with a radius parameter. A constant $\tau_c$ is considered above the critical temperature. With this parametrization and a phenomenological temperature-dependent angular velocity $\Om(T)$, we obtained scaled conductivity and ratio of shear viscosity to entropy density as a function of temperature.
Unlike a magnetic field, rotation does not distinguish positively and negatively charged particles; as a result, the Hall-like transport component induced by rotation makes a significant contribution to transportation. Moreover, an appreciable amount of anisotropy in the transport coefficients is noticed around kinetic freeze-out ($T\approx 100$ MeV), which can, in principle, affect the particle spectra. 
\section{Acknowledgement}
AD gratefully acknowledges the Ministry of Education (MoE), Government of India. KG acknowledges the financial support from the Prime Minister’s Research Fellowship (PMRF), Government of India. KG and RS acknowledge the DAE-DST, Government of India funding under the mega-science project “Indian participation in the ALICE experiment at CERN” bearing Project No. SR/MF/PS-02/2021-IITI(E-37123). SG acknowledges the DAE-BRNS project with Grant Nos. 57/14/01/2024-BRNS/313, Government of India.   
\bibliographystyle{apsrev4-2}
\bibliography{reference}

\begin{thebibliography}{81}%
\makeatletter
\providecommand \@ifxundefined [1]{%
 \@ifx{#1\undefined}
}%
\providecommand \@ifnum [1]{%
 \ifnum #1\expandafter \@firstoftwo
 \else \expandafter \@secondoftwo
 \fi
}%
\providecommand \@ifx [1]{%
 \ifx #1\expandafter \@firstoftwo
 \else \expandafter \@secondoftwo
 \fi
}%
\providecommand \natexlab [1]{#1}%
\providecommand \enquote  [1]{``#1''}%
\providecommand \bibnamefont  [1]{#1}%
\providecommand \bibfnamefont [1]{#1}%
\providecommand \citenamefont [1]{#1}%
\providecommand \href@noop [0]{\@secondoftwo}%
\providecommand \href [0]{\begingroup \@sanitize@url \@href}%
\providecommand \@href[1]{\@@startlink{#1}\@@href}%
\providecommand \@@href[1]{\endgroup#1\@@endlink}%
\providecommand \@sanitize@url [0]{\catcode `\\12\catcode `\$12\catcode
  `\&12\catcode `\#12\catcode `\^12\catcode `\_12\catcode `\%12\relax}%
\providecommand \@@startlink[1]{}%
\providecommand \@@endlink[0]{}%
\providecommand \url  [0]{\begingroup\@sanitize@url \@url }%
\providecommand \@url [1]{\endgroup\@href {#1}{\urlprefix }}%
\providecommand \urlprefix  [0]{URL }%
\providecommand \Eprint [0]{\href }%
\providecommand \doibase [0]{https://doi.org/}%
\providecommand \selectlanguage [0]{\@gobble}%
\providecommand \bibinfo  [0]{\@secondoftwo}%
\providecommand \bibfield  [0]{\@secondoftwo}%
\providecommand \translation [1]{[#1]}%
\providecommand \BibitemOpen [0]{}%
\providecommand \bibitemStop [0]{}%
\providecommand \bibitemNoStop [0]{.\EOS\space}%
\providecommand \EOS [0]{\spacefactor3000\relax}%
\providecommand \BibitemShut  [1]{\csname bibitem#1\endcsname}%
\let\auto@bib@innerbib\@empty
%</preamble>
\bibitem [{\citenamefont {Shuryak}(2004)}]{Shuryak:2003xe}%
  \BibitemOpen
  \bibfield  {author} {\bibinfo {author} {\bibfnamefont {E.}~\bibnamefont
  {Shuryak}},\ }\href {https://doi.org/10.1016/j.ppnp.2004.02.025} {\bibfield
  {journal} {\bibinfo  {journal} {Prog. Part. Nucl. Phys.}\ }\textbf {\bibinfo
  {volume} {53}},\ \bibinfo {pages} {273} (\bibinfo {year} {2004})},\ \Eprint
  {https://arxiv.org/abs/hep-ph/0312227} {arXiv:hep-ph/0312227} \BibitemShut
  {NoStop}%
\bibitem [{\citenamefont {Heinz}(2004)}]{Heinz:2004qz}%
  \BibitemOpen
  \bibfield  {author} {\bibinfo {author} {\bibfnamefont {U.~W.}\ \bibnamefont
  {Heinz}},\ }in\ \href@noop {} {\emph {\bibinfo {booktitle} {{2nd CERN-CLAF
  School of High Energy Physics}}}}\ (\bibinfo {year} {2004})\ pp.\ \bibinfo
  {pages} {165--238},\ \Eprint {https://arxiv.org/abs/hep-ph/0407360}
  {arXiv:hep-ph/0407360} \BibitemShut {NoStop}%
\bibitem [{\citenamefont {Skokov}\ \emph {et~al.}(2009)\citenamefont {Skokov},
  \citenamefont {Illarionov},\ and\ \citenamefont {Toneev}}]{Skokov:2009qp}%
  \BibitemOpen
  \bibfield  {author} {\bibinfo {author} {\bibfnamefont {V.}~\bibnamefont
  {Skokov}}, \bibinfo {author} {\bibfnamefont {A.~Y.}\ \bibnamefont
  {Illarionov}},\ and\ \bibinfo {author} {\bibfnamefont {V.}~\bibnamefont
  {Toneev}},\ }\href {https://doi.org/10.1142/S0217751X09047570} {\bibfield
  {journal} {\bibinfo  {journal} {Int. J. Mod. Phys. A}\ }\textbf {\bibinfo
  {volume} {24}},\ \bibinfo {pages} {5925} (\bibinfo {year} {2009})},\ \Eprint
  {https://arxiv.org/abs/0907.1396} {arXiv:0907.1396 [nucl-th]} \BibitemShut
  {NoStop}%
\bibitem [{\citenamefont {Bzdak}\ and\ \citenamefont
  {Skokov}(2012)}]{Bzdak:2011yy}%
  \BibitemOpen
  \bibfield  {author} {\bibinfo {author} {\bibfnamefont {A.}~\bibnamefont
  {Bzdak}}\ and\ \bibinfo {author} {\bibfnamefont {V.}~\bibnamefont {Skokov}},\
  }\href {https://doi.org/10.1016/j.physletb.2012.02.065} {\bibfield  {journal}
  {\bibinfo  {journal} {Phys. Lett. B}\ }\textbf {\bibinfo {volume} {710}},\
  \bibinfo {pages} {171} (\bibinfo {year} {2012})},\ \Eprint
  {https://arxiv.org/abs/1111.1949} {arXiv:1111.1949 [hep-ph]} \BibitemShut
  {NoStop}%
\bibitem [{\citenamefont {Abdallah}\ \emph {et~al.}(2023)\citenamefont
  {Abdallah} \emph {et~al.}}]{STAR:2022fan}%
  \BibitemOpen
  \bibfield  {author} {\bibinfo {author} {\bibfnamefont {M.~S.}\ \bibnamefont
  {Abdallah}} \emph {et~al.} (\bibinfo {collaboration} {STAR}),\ }\href
  {https://doi.org/10.1038/s41586-022-05557-5} {\bibfield  {journal} {\bibinfo
  {journal} {Nature}\ }\textbf {\bibinfo {volume} {614}},\ \bibinfo {pages}
  {244} (\bibinfo {year} {2023})},\ \Eprint {https://arxiv.org/abs/2204.02302}
  {arXiv:2204.02302 [hep-ph]} \BibitemShut {NoStop}%
\bibitem [{\citenamefont {{Einstein}}\ and\ \citenamefont {{de
  Haas}}(1915)}]{1915DPhyG17152E}%
  \BibitemOpen
  \bibfield  {author} {\bibinfo {author} {\bibfnamefont {A.}~\bibnamefont
  {{Einstein}}}\ and\ \bibinfo {author} {\bibfnamefont {W.~J.}\ \bibnamefont
  {{de Haas}}},\ }\href@noop {} {\bibfield  {journal} {\bibinfo  {journal}
  {Deutsche Physikalische Gesellschaft}\ }\textbf {\bibinfo {volume} {17}},\
  \bibinfo {pages} {152} (\bibinfo {year} {1915})}\BibitemShut {NoStop}%
\bibitem [{\citenamefont {{Richardson}}(1908)}]{1908Richardson}%
  \BibitemOpen
  \bibfield  {author} {\bibinfo {author} {\bibfnamefont {O.~W.}\ \bibnamefont
  {{Richardson}}},\ }\href {https://doi.org/10.1103/PhysRevSeriesI.26.248}
  {\bibfield  {journal} {\bibinfo  {journal} {Physical Review Series I}\
  }\textbf {\bibinfo {volume} {26}},\ \bibinfo {pages} {248} (\bibinfo {year}
  {1908})}\BibitemShut {NoStop}%
\bibitem [{\citenamefont {Sivardiere}(1983)}]{J_Sivardiere_1983}%
  \BibitemOpen
  \bibfield  {author} {\bibinfo {author} {\bibfnamefont {J.}~\bibnamefont
  {Sivardiere}},\ }\href {https://doi.org/10.1088/0143-0807/4/3/007} {\bibfield
   {journal} {\bibinfo  {journal} {European Journal of Physics}\ }\textbf
  {\bibinfo {volume} {4}},\ \bibinfo {pages} {162} (\bibinfo {year}
  {1983})}\BibitemShut {NoStop}%
\bibitem [{\citenamefont {Johnson}(2000)}]{Johnson2000-px}%
  \BibitemOpen
  \bibfield  {author} {\bibinfo {author} {\bibfnamefont {B.~L.}\ \bibnamefont
  {Johnson}},\ }\href@noop {} {\bibfield  {journal} {\bibinfo  {journal} {Am.
  J. Phys.}\ }\textbf {\bibinfo {volume} {68}},\ \bibinfo {pages} {649}
  (\bibinfo {year} {2000})}\BibitemShut {NoStop}%
\bibitem [{\citenamefont {Sakurai}(1980)}]{Sakurai1980}%
  \BibitemOpen
  \bibfield  {author} {\bibinfo {author} {\bibfnamefont {J.~J.}\ \bibnamefont
  {Sakurai}},\ }\href {https://doi.org/10.1103/PhysRevD.21.2993} {\bibfield
  {journal} {\bibinfo  {journal} {Phys. Rev. D}\ }\textbf {\bibinfo {volume}
  {21}},\ \bibinfo {pages} {2993} (\bibinfo {year} {1980})}\BibitemShut
  {NoStop}%
\bibitem [{\citenamefont {Brandão}\ \emph {et~al.}(2015)\citenamefont
  {Brandão}, \citenamefont {Moraes}, \citenamefont {Cunha}, \citenamefont
  {Lima},\ and\ \citenamefont {Filgueiras}}]{BRANDAO201555}%
  \BibitemOpen
  \bibfield  {author} {\bibinfo {author} {\bibfnamefont {J.~E.}\ \bibnamefont
  {Brandão}}, \bibinfo {author} {\bibfnamefont {F.}~\bibnamefont {Moraes}},
  \bibinfo {author} {\bibfnamefont {M.}~\bibnamefont {Cunha}}, \bibinfo
  {author} {\bibfnamefont {J.~R.}\ \bibnamefont {Lima}},\ and\ \bibinfo
  {author} {\bibfnamefont {C.}~\bibnamefont {Filgueiras}},\ }\href
  {https://doi.org/https://doi.org/10.1016/j.rinp.2015.02.003} {\bibfield
  {journal} {\bibinfo  {journal} {Results in Physics}\ }\textbf {\bibinfo
  {volume} {5}},\ \bibinfo {pages} {55} (\bibinfo {year} {2015})}\BibitemShut
  {NoStop}%
\bibitem [{\citenamefont {Chen}\ \emph {et~al.}(2016)\citenamefont {Chen},
  \citenamefont {Fukushima}, \citenamefont {Huang},\ and\ \citenamefont
  {Mameda}}]{Chen:2015hfc}%
  \BibitemOpen
  \bibfield  {author} {\bibinfo {author} {\bibfnamefont {H.-L.}\ \bibnamefont
  {Chen}}, \bibinfo {author} {\bibfnamefont {K.}~\bibnamefont {Fukushima}},
  \bibinfo {author} {\bibfnamefont {X.-G.}\ \bibnamefont {Huang}},\ and\
  \bibinfo {author} {\bibfnamefont {K.}~\bibnamefont {Mameda}},\ }\href
  {https://doi.org/10.1103/PhysRevD.93.104052} {\bibfield  {journal} {\bibinfo
  {journal} {Phys. Rev. D}\ }\textbf {\bibinfo {volume} {93}},\ \bibinfo
  {pages} {104052} (\bibinfo {year} {2016})},\ \Eprint
  {https://arxiv.org/abs/1512.08974} {arXiv:1512.08974 [hep-ph]} \BibitemShut
  {NoStop}%
\bibitem [{\citenamefont {Mameda}\ and\ \citenamefont
  {Yamamoto}(2016)}]{Mameda:2015ria}%
  \BibitemOpen
  \bibfield  {author} {\bibinfo {author} {\bibfnamefont {K.}~\bibnamefont
  {Mameda}}\ and\ \bibinfo {author} {\bibfnamefont {A.}~\bibnamefont
  {Yamamoto}},\ }\href {https://doi.org/10.1093/ptep/ptw128} {\bibfield
  {journal} {\bibinfo  {journal} {PTEP}\ }\textbf {\bibinfo {volume} {2016}},\
  \bibinfo {pages} {093B05} (\bibinfo {year} {2016})},\ \Eprint
  {https://arxiv.org/abs/1504.05826} {arXiv:1504.05826 [hep-th]} \BibitemShut
  {NoStop}%
\bibitem [{\citenamefont {Ghosh}\ \emph {et~al.}(2020)\citenamefont {Ghosh},
  \citenamefont {Bandyopadhyay}, \citenamefont {Farias}, \citenamefont {Dey},\
  and\ \citenamefont {Krein}}]{Ghosh:2019ubc}%
  \BibitemOpen
  \bibfield  {author} {\bibinfo {author} {\bibfnamefont {S.}~\bibnamefont
  {Ghosh}}, \bibinfo {author} {\bibfnamefont {A.}~\bibnamefont
  {Bandyopadhyay}}, \bibinfo {author} {\bibfnamefont {R.~L.~S.}\ \bibnamefont
  {Farias}}, \bibinfo {author} {\bibfnamefont {J.}~\bibnamefont {Dey}},\ and\
  \bibinfo {author} {\bibfnamefont {G.~a.}\ \bibnamefont {Krein}},\ }\href
  {https://doi.org/10.1103/PhysRevD.102.114015} {\bibfield  {journal} {\bibinfo
   {journal} {Phys. Rev. D}\ }\textbf {\bibinfo {volume} {102}},\ \bibinfo
  {pages} {114015} (\bibinfo {year} {2020})},\ \Eprint
  {https://arxiv.org/abs/1911.10005} {arXiv:1911.10005 [hep-ph]} \BibitemShut
  {NoStop}%
\bibitem [{\citenamefont {Dey}\ \emph {et~al.}(2022)\citenamefont {Dey},
  \citenamefont {Samanta}, \citenamefont {Ghosh},\ and\ \citenamefont
  {Satapathy}}]{Dey:2020awu}%
  \BibitemOpen
  \bibfield  {author} {\bibinfo {author} {\bibfnamefont {J.}~\bibnamefont
  {Dey}}, \bibinfo {author} {\bibfnamefont {S.}~\bibnamefont {Samanta}},
  \bibinfo {author} {\bibfnamefont {S.}~\bibnamefont {Ghosh}},\ and\ \bibinfo
  {author} {\bibfnamefont {S.}~\bibnamefont {Satapathy}},\ }\href
  {https://doi.org/10.1103/PhysRevC.106.044914} {\bibfield  {journal} {\bibinfo
   {journal} {Phys. Rev. C}\ }\textbf {\bibinfo {volume} {106}},\ \bibinfo
  {pages} {044914} (\bibinfo {year} {2022})},\ \Eprint
  {https://arxiv.org/abs/2002.04434} {arXiv:2002.04434 [nucl-th]} \BibitemShut
  {NoStop}%
\bibitem [{\citenamefont {Kalikotay}\ \emph {et~al.}(2020)\citenamefont
  {Kalikotay}, \citenamefont {Ghosh}, \citenamefont {Chaudhuri}, \citenamefont
  {Roy},\ and\ \citenamefont {Sarkar}}]{Kalikotay:2020snc}%
  \BibitemOpen
  \bibfield  {author} {\bibinfo {author} {\bibfnamefont {P.}~\bibnamefont
  {Kalikotay}}, \bibinfo {author} {\bibfnamefont {S.}~\bibnamefont {Ghosh}},
  \bibinfo {author} {\bibfnamefont {N.}~\bibnamefont {Chaudhuri}}, \bibinfo
  {author} {\bibfnamefont {P.}~\bibnamefont {Roy}},\ and\ \bibinfo {author}
  {\bibfnamefont {S.}~\bibnamefont {Sarkar}},\ }\href
  {https://doi.org/10.1103/PhysRevD.102.076007} {\bibfield  {journal} {\bibinfo
   {journal} {Phys. Rev. D}\ }\textbf {\bibinfo {volume} {102}},\ \bibinfo
  {pages} {076007} (\bibinfo {year} {2020})},\ \Eprint
  {https://arxiv.org/abs/2009.10493} {arXiv:2009.10493 [hep-ph]} \BibitemShut
  {NoStop}%
\bibitem [{\citenamefont {Dey}\ \emph {et~al.}(2023)\citenamefont {Dey},
  \citenamefont {Bandyopadhyay}, \citenamefont {Gupta}, \citenamefont
  {Pujari},\ and\ \citenamefont {Ghosh}}]{Dey:2021fbo}%
  \BibitemOpen
  \bibfield  {author} {\bibinfo {author} {\bibfnamefont {J.}~\bibnamefont
  {Dey}}, \bibinfo {author} {\bibfnamefont {A.}~\bibnamefont {Bandyopadhyay}},
  \bibinfo {author} {\bibfnamefont {A.}~\bibnamefont {Gupta}}, \bibinfo
  {author} {\bibfnamefont {N.}~\bibnamefont {Pujari}},\ and\ \bibinfo {author}
  {\bibfnamefont {S.}~\bibnamefont {Ghosh}},\ }\href
  {https://doi.org/10.1016/j.nuclphysa.2023.122654} {\bibfield  {journal}
  {\bibinfo  {journal} {Nucl. Phys. A}\ }\textbf {\bibinfo {volume} {1034}},\
  \bibinfo {pages} {122654} (\bibinfo {year} {2023})},\ \Eprint
  {https://arxiv.org/abs/2103.15364} {arXiv:2103.15364 [hep-ph]} \BibitemShut
  {NoStop}%
\bibitem [{\citenamefont {Satapathy}\ \emph {et~al.}(2021)\citenamefont
  {Satapathy}, \citenamefont {Ghosh},\ and\ \citenamefont
  {Ghosh}}]{Satapathy:2021cjp}%
  \BibitemOpen
  \bibfield  {author} {\bibinfo {author} {\bibfnamefont {S.}~\bibnamefont
  {Satapathy}}, \bibinfo {author} {\bibfnamefont {S.}~\bibnamefont {Ghosh}},\
  and\ \bibinfo {author} {\bibfnamefont {S.}~\bibnamefont {Ghosh}},\ }\href
  {https://doi.org/10.1103/PhysRevD.104.056030} {\bibfield  {journal} {\bibinfo
   {journal} {Phys. Rev. D}\ }\textbf {\bibinfo {volume} {104}},\ \bibinfo
  {pages} {056030} (\bibinfo {year} {2021})},\ \Eprint
  {https://arxiv.org/abs/2104.03917} {arXiv:2104.03917 [hep-ph]} \BibitemShut
  {NoStop}%
\bibitem [{\citenamefont {Das}\ \emph {et~al.}(2019)\citenamefont {Das},
  \citenamefont {Mishra},\ and\ \citenamefont {Mohapatra}}]{Das:2019wjg}%
  \BibitemOpen
  \bibfield  {author} {\bibinfo {author} {\bibfnamefont {A.}~\bibnamefont
  {Das}}, \bibinfo {author} {\bibfnamefont {H.}~\bibnamefont {Mishra}},\ and\
  \bibinfo {author} {\bibfnamefont {R.~K.}\ \bibnamefont {Mohapatra}},\ }\href
  {https://doi.org/10.1103/PhysRevD.99.094031} {\bibfield  {journal} {\bibinfo
  {journal} {Phys. Rev. D}\ }\textbf {\bibinfo {volume} {99}},\ \bibinfo
  {pages} {094031} (\bibinfo {year} {2019})},\ \Eprint
  {https://arxiv.org/abs/1903.03938} {arXiv:1903.03938 [hep-ph]} \BibitemShut
  {NoStop}%
\bibitem [{\citenamefont {Das}\ \emph {et~al.}(2020)\citenamefont {Das},
  \citenamefont {Mishra},\ and\ \citenamefont {Mohapatra}}]{Das:2019ppb}%
  \BibitemOpen
  \bibfield  {author} {\bibinfo {author} {\bibfnamefont {A.}~\bibnamefont
  {Das}}, \bibinfo {author} {\bibfnamefont {H.}~\bibnamefont {Mishra}},\ and\
  \bibinfo {author} {\bibfnamefont {R.~K.}\ \bibnamefont {Mohapatra}},\ }\href
  {https://doi.org/10.1103/PhysRevD.101.034027} {\bibfield  {journal} {\bibinfo
   {journal} {Phys. Rev. D}\ }\textbf {\bibinfo {volume} {101}},\ \bibinfo
  {pages} {034027} (\bibinfo {year} {2020})},\ \Eprint
  {https://arxiv.org/abs/1907.05298} {arXiv:1907.05298 [hep-ph]} \BibitemShut
  {NoStop}%
\bibitem [{\citenamefont {Chatterjee}\ \emph {et~al.}(2021)\citenamefont
  {Chatterjee}, \citenamefont {Rath}, \citenamefont {Sarwar},\ and\
  \citenamefont {Sahoo}}]{Chatterjee:2019nld}%
  \BibitemOpen
  \bibfield  {author} {\bibinfo {author} {\bibfnamefont {B.}~\bibnamefont
  {Chatterjee}}, \bibinfo {author} {\bibfnamefont {R.}~\bibnamefont {Rath}},
  \bibinfo {author} {\bibfnamefont {G.}~\bibnamefont {Sarwar}},\ and\ \bibinfo
  {author} {\bibfnamefont {R.}~\bibnamefont {Sahoo}},\ }\href
  {https://doi.org/10.1140/epja/s10050-021-00348-4} {\bibfield  {journal}
  {\bibinfo  {journal} {Eur. Phys. J. A}\ }\textbf {\bibinfo {volume} {57}},\
  \bibinfo {pages} {45} (\bibinfo {year} {2021})},\ \Eprint
  {https://arxiv.org/abs/1908.01121} {arXiv:1908.01121 [hep-ph]} \BibitemShut
  {NoStop}%
\bibitem [{\citenamefont {Hattori}\ and\ \citenamefont
  {Satow}(2016)}]{Hattori:2016cnt}%
  \BibitemOpen
  \bibfield  {author} {\bibinfo {author} {\bibfnamefont {K.}~\bibnamefont
  {Hattori}}\ and\ \bibinfo {author} {\bibfnamefont {D.}~\bibnamefont
  {Satow}},\ }\href {https://doi.org/10.1103/PhysRevD.94.114032} {\bibfield
  {journal} {\bibinfo  {journal} {Phys. Rev. D}\ }\textbf {\bibinfo {volume}
  {94}},\ \bibinfo {pages} {114032} (\bibinfo {year} {2016})},\ \Eprint
  {https://arxiv.org/abs/1610.06818} {arXiv:1610.06818 [hep-ph]} \BibitemShut
  {NoStop}%
\bibitem [{\citenamefont {Hattori}\ \emph {et~al.}(2017)\citenamefont
  {Hattori}, \citenamefont {Li}, \citenamefont {Satow},\ and\ \citenamefont
  {Yee}}]{Hattori:2016lqx}%
  \BibitemOpen
  \bibfield  {author} {\bibinfo {author} {\bibfnamefont {K.}~\bibnamefont
  {Hattori}}, \bibinfo {author} {\bibfnamefont {S.}~\bibnamefont {Li}},
  \bibinfo {author} {\bibfnamefont {D.}~\bibnamefont {Satow}},\ and\ \bibinfo
  {author} {\bibfnamefont {H.-U.}\ \bibnamefont {Yee}},\ }\href
  {https://doi.org/10.1103/PhysRevD.95.076008} {\bibfield  {journal} {\bibinfo
  {journal} {Phys. Rev. D}\ }\textbf {\bibinfo {volume} {95}},\ \bibinfo
  {pages} {076008} (\bibinfo {year} {2017})},\ \Eprint
  {https://arxiv.org/abs/1610.06839} {arXiv:1610.06839 [hep-ph]} \BibitemShut
  {NoStop}%
\bibitem [{\citenamefont {Satapathy}\ \emph {et~al.}(2022)\citenamefont
  {Satapathy}, \citenamefont {Ghosh},\ and\ \citenamefont
  {Ghosh}}]{Satapathy:2021wex}%
  \BibitemOpen
  \bibfield  {author} {\bibinfo {author} {\bibfnamefont {S.}~\bibnamefont
  {Satapathy}}, \bibinfo {author} {\bibfnamefont {S.}~\bibnamefont {Ghosh}},\
  and\ \bibinfo {author} {\bibfnamefont {S.}~\bibnamefont {Ghosh}},\ }\href
  {https://doi.org/10.1103/PhysRevD.106.036006} {\bibfield  {journal} {\bibinfo
   {journal} {Phys. Rev. D}\ }\textbf {\bibinfo {volume} {106}},\ \bibinfo
  {pages} {036006} (\bibinfo {year} {2022})},\ \Eprint
  {https://arxiv.org/abs/2112.08236} {arXiv:2112.08236 [hep-ph]} \BibitemShut
  {NoStop}%
\bibitem [{\citenamefont {Andersen}(2021)}]{Andersen:2021lnk}%
  \BibitemOpen
  \bibfield  {author} {\bibinfo {author} {\bibfnamefont {J.~O.}\ \bibnamefont
  {Andersen}},\ }\href {https://doi.org/10.1140/epja/s10050-021-00491-y}
  {\bibfield  {journal} {\bibinfo  {journal} {Eur. Phys. J. A}\ }\textbf
  {\bibinfo {volume} {57}},\ \bibinfo {pages} {189} (\bibinfo {year} {2021})},\
  \Eprint {https://arxiv.org/abs/2102.13165} {arXiv:2102.13165 [hep-ph]}
  \BibitemShut {NoStop}%
\bibitem [{\citenamefont {Li}\ and\ \citenamefont {Wang}(2020)}]{Li:2020dwr}%
  \BibitemOpen
  \bibfield  {author} {\bibinfo {author} {\bibfnamefont {W.}~\bibnamefont
  {Li}}\ and\ \bibinfo {author} {\bibfnamefont {G.}~\bibnamefont {Wang}},\
  }\href {https://doi.org/10.1146/annurev-nucl-030220-065203} {\bibfield
  {journal} {\bibinfo  {journal} {Ann. Rev. Nucl. Part. Sci.}\ }\textbf
  {\bibinfo {volume} {70}},\ \bibinfo {pages} {293} (\bibinfo {year} {2020})},\
  \Eprint {https://arxiv.org/abs/2002.10397} {arXiv:2002.10397 [nucl-ex]}
  \BibitemShut {NoStop}%
\bibitem [{\citenamefont {Goswami}\ \emph {et~al.}(2024)\citenamefont
  {Goswami}, \citenamefont {Sahu}, \citenamefont {Dey}, \citenamefont {Sahoo},\
  and\ \citenamefont {Stock}}]{Goswami:2023eol}%
  \BibitemOpen
  \bibfield  {author} {\bibinfo {author} {\bibfnamefont {K.}~\bibnamefont
  {Goswami}}, \bibinfo {author} {\bibfnamefont {D.}~\bibnamefont {Sahu}},
  \bibinfo {author} {\bibfnamefont {J.}~\bibnamefont {Dey}}, \bibinfo {author}
  {\bibfnamefont {R.}~\bibnamefont {Sahoo}},\ and\ \bibinfo {author}
  {\bibfnamefont {R.}~\bibnamefont {Stock}},\ }\href
  {https://doi.org/10.1103/PhysRevD.109.074012} {\bibfield  {journal} {\bibinfo
   {journal} {Phys. Rev. D}\ }\textbf {\bibinfo {volume} {109}},\ \bibinfo
  {pages} {074012} (\bibinfo {year} {2024})},\ \Eprint
  {https://arxiv.org/abs/2310.02711} {arXiv:2310.02711 [hep-ph]} \BibitemShut
  {NoStop}%
\bibitem [{\citenamefont {Sahoo}\ \emph {et~al.}(2023)\citenamefont {Sahoo},
  \citenamefont {Pradhan}, \citenamefont {Sahu},\ and\ \citenamefont
  {Sahoo}}]{Sahoo:2023vkw}%
  \BibitemOpen
  \bibfield  {author} {\bibinfo {author} {\bibfnamefont {B.}~\bibnamefont
  {Sahoo}}, \bibinfo {author} {\bibfnamefont {K.~K.}\ \bibnamefont {Pradhan}},
  \bibinfo {author} {\bibfnamefont {D.}~\bibnamefont {Sahu}},\ and\ \bibinfo
  {author} {\bibfnamefont {R.}~\bibnamefont {Sahoo}},\ }\href
  {https://doi.org/10.1103/PhysRevD.108.074028} {\bibfield  {journal} {\bibinfo
   {journal} {Phys. Rev. D}\ }\textbf {\bibinfo {volume} {108}},\ \bibinfo
  {pages} {074028} (\bibinfo {year} {2023})},\ \Eprint
  {https://arxiv.org/abs/2306.03477} {arXiv:2306.03477 [hep-ph]} \BibitemShut
  {NoStop}%
\bibitem [{\citenamefont {Becattini}\ \emph {et~al.}(2021)\citenamefont
  {Becattini}, \citenamefont {Liao},\ and\ \citenamefont
  {Lisa}}]{Becattini:2021lfq}%
  \BibitemOpen
  \bibfield  {author} {\bibinfo {author} {\bibfnamefont {F.}~\bibnamefont
  {Becattini}}, \bibinfo {author} {\bibfnamefont {J.}~\bibnamefont {Liao}},\
  and\ \bibinfo {author} {\bibfnamefont {M.}~\bibnamefont {Lisa}},\ }\href
  {https://doi.org/10.1007/978-3-030-71427-7_1} {\bibfield  {journal} {\bibinfo
   {journal} {Lect. Notes Phys.}\ }\textbf {\bibinfo {volume} {987}},\ \bibinfo
  {pages} {1} (\bibinfo {year} {2021})},\ \Eprint
  {https://arxiv.org/abs/2102.00933} {arXiv:2102.00933 [nucl-th]} \BibitemShut
  {NoStop}%
\bibitem [{\citenamefont {Wang}\ \emph {et~al.}(2019)\citenamefont {Wang},
  \citenamefont {Wei}, \citenamefont {Li},\ and\ \citenamefont
  {Huang}}]{Wang:2018sur}%
  \BibitemOpen
  \bibfield  {author} {\bibinfo {author} {\bibfnamefont {X.}~\bibnamefont
  {Wang}}, \bibinfo {author} {\bibfnamefont {M.}~\bibnamefont {Wei}}, \bibinfo
  {author} {\bibfnamefont {Z.}~\bibnamefont {Li}},\ and\ \bibinfo {author}
  {\bibfnamefont {M.}~\bibnamefont {Huang}},\ }\href
  {https://doi.org/10.1103/PhysRevD.99.016018} {\bibfield  {journal} {\bibinfo
  {journal} {Phys. Rev. D}\ }\textbf {\bibinfo {volume} {99}},\ \bibinfo
  {pages} {016018} (\bibinfo {year} {2019})},\ \Eprint
  {https://arxiv.org/abs/1808.01931} {arXiv:1808.01931 [hep-ph]} \BibitemShut
  {NoStop}%
\bibitem [{\citenamefont {Fujimoto}\ \emph {et~al.}(2021)\citenamefont
  {Fujimoto}, \citenamefont {Fukushima},\ and\ \citenamefont
  {Hidaka}}]{Fujimoto:2021xix}%
  \BibitemOpen
  \bibfield  {author} {\bibinfo {author} {\bibfnamefont {Y.}~\bibnamefont
  {Fujimoto}}, \bibinfo {author} {\bibfnamefont {K.}~\bibnamefont
  {Fukushima}},\ and\ \bibinfo {author} {\bibfnamefont {Y.}~\bibnamefont
  {Hidaka}},\ }\href {https://doi.org/10.1016/j.physletb.2021.136184}
  {\bibfield  {journal} {\bibinfo  {journal} {Phys. Lett. B}\ }\textbf
  {\bibinfo {volume} {816}},\ \bibinfo {pages} {136184} (\bibinfo {year}
  {2021})},\ \Eprint {https://arxiv.org/abs/2101.09173} {arXiv:2101.09173
  [hep-ph]} \BibitemShut {NoStop}%
\bibitem [{\citenamefont {Pradhan}\ \emph {et~al.}(2024)\citenamefont
  {Pradhan}, \citenamefont {Sahoo}, \citenamefont {Sahu},\ and\ \citenamefont
  {Sahoo}}]{Pradhan:2023rvf}%
  \BibitemOpen
  \bibfield  {author} {\bibinfo {author} {\bibfnamefont {K.~K.}\ \bibnamefont
  {Pradhan}}, \bibinfo {author} {\bibfnamefont {B.}~\bibnamefont {Sahoo}},
  \bibinfo {author} {\bibfnamefont {D.}~\bibnamefont {Sahu}},\ and\ \bibinfo
  {author} {\bibfnamefont {R.}~\bibnamefont {Sahoo}},\ }\href
  {https://doi.org/10.1140/epjc/s10052-024-13283-7} {\bibfield  {journal}
  {\bibinfo  {journal} {Eur. Phys. J. C}\ }\textbf {\bibinfo {volume} {84}},\
  \bibinfo {pages} {936} (\bibinfo {year} {2024})},\ \Eprint
  {https://arxiv.org/abs/2304.05190} {arXiv:2304.05190 [hep-ph]} \BibitemShut
  {NoStop}%
\bibitem [{\citenamefont {Braguta}\ \emph {et~al.}(2021)\citenamefont
  {Braguta}, \citenamefont {Kotov}, \citenamefont {Kuznedelev},\ and\
  \citenamefont {Roenko}}]{Braguta:2021jgn}%
  \BibitemOpen
  \bibfield  {author} {\bibinfo {author} {\bibfnamefont {V.~V.}\ \bibnamefont
  {Braguta}}, \bibinfo {author} {\bibfnamefont {A.~Y.}\ \bibnamefont {Kotov}},
  \bibinfo {author} {\bibfnamefont {D.~D.}\ \bibnamefont {Kuznedelev}},\ and\
  \bibinfo {author} {\bibfnamefont {A.~A.}\ \bibnamefont {Roenko}},\ }\href
  {https://doi.org/10.1103/PhysRevD.103.094515} {\bibfield  {journal} {\bibinfo
   {journal} {Phys. Rev. D}\ }\textbf {\bibinfo {volume} {103}},\ \bibinfo
  {pages} {094515} (\bibinfo {year} {2021})},\ \Eprint
  {https://arxiv.org/abs/2102.05084} {arXiv:2102.05084 [hep-lat]} \BibitemShut
  {NoStop}%
\bibitem [{\citenamefont {Nambu}\ and\ \citenamefont
  {Jona-Lasinio}(1961{\natexlab{a}})}]{Nambu:1961fr}%
  \BibitemOpen
  \bibfield  {author} {\bibinfo {author} {\bibfnamefont {Y.}~\bibnamefont
  {Nambu}}\ and\ \bibinfo {author} {\bibfnamefont {G.}~\bibnamefont
  {Jona-Lasinio}},\ }\href {https://doi.org/10.1103/PhysRev.124.246} {\bibfield
   {journal} {\bibinfo  {journal} {Phys. Rev.}\ }\textbf {\bibinfo {volume}
  {124}},\ \bibinfo {pages} {246} (\bibinfo {year}
  {1961}{\natexlab{a}})}\BibitemShut {NoStop}%
\bibitem [{\citenamefont {Nambu}\ and\ \citenamefont
  {Jona-Lasinio}(1961{\natexlab{b}})}]{Nambu:1961tp}%
  \BibitemOpen
  \bibfield  {author} {\bibinfo {author} {\bibfnamefont {Y.}~\bibnamefont
  {Nambu}}\ and\ \bibinfo {author} {\bibfnamefont {G.}~\bibnamefont
  {Jona-Lasinio}},\ }\href {https://doi.org/10.1103/PhysRev.122.345} {\bibfield
   {journal} {\bibinfo  {journal} {Phys. Rev.}\ }\textbf {\bibinfo {volume}
  {122}},\ \bibinfo {pages} {345} (\bibinfo {year}
  {1961}{\natexlab{b}})}\BibitemShut {NoStop}%
\bibitem [{\citenamefont {Hatsuda}\ and\ \citenamefont
  {Kunihiro}(1994)}]{Hatsuda:1994pi}%
  \BibitemOpen
  \bibfield  {author} {\bibinfo {author} {\bibfnamefont {T.}~\bibnamefont
  {Hatsuda}}\ and\ \bibinfo {author} {\bibfnamefont {T.}~\bibnamefont
  {Kunihiro}},\ }\href {https://doi.org/10.1016/0370-1573(94)90022-1}
  {\bibfield  {journal} {\bibinfo  {journal} {Phys. Rept.}\ }\textbf {\bibinfo
  {volume} {247}},\ \bibinfo {pages} {221} (\bibinfo {year} {1994})},\ \Eprint
  {https://arxiv.org/abs/hep-ph/9401310} {arXiv:hep-ph/9401310} \BibitemShut
  {NoStop}%
\bibitem [{\citenamefont {Klevansky}(1992)}]{Klevansky:1992qe}%
  \BibitemOpen
  \bibfield  {author} {\bibinfo {author} {\bibfnamefont {S.~P.}\ \bibnamefont
  {Klevansky}},\ }\href {https://doi.org/10.1103/RevModPhys.64.649} {\bibfield
  {journal} {\bibinfo  {journal} {Rev. Mod. Phys.}\ }\textbf {\bibinfo {volume}
  {64}},\ \bibinfo {pages} {649} (\bibinfo {year} {1992})}\BibitemShut
  {NoStop}%
\bibitem [{\citenamefont {Kovtun}\ \emph {et~al.}(2005)\citenamefont {Kovtun},
  \citenamefont {Son},\ and\ \citenamefont {Starinets}}]{Kovtun:2004de}%
  \BibitemOpen
  \bibfield  {author} {\bibinfo {author} {\bibfnamefont {P.}~\bibnamefont
  {Kovtun}}, \bibinfo {author} {\bibfnamefont {D.~T.}\ \bibnamefont {Son}},\
  and\ \bibinfo {author} {\bibfnamefont {A.~O.}\ \bibnamefont {Starinets}},\
  }\href {https://doi.org/10.1103/PhysRevLett.94.111601} {\bibfield  {journal}
  {\bibinfo  {journal} {Phys. Rev. Lett.}\ }\textbf {\bibinfo {volume} {94}},\
  \bibinfo {pages} {111601} (\bibinfo {year} {2005})},\ \Eprint
  {https://arxiv.org/abs/hep-th/0405231} {arXiv:hep-th/0405231} \BibitemShut
  {NoStop}%
\bibitem [{\citenamefont {Yang}\ and\ \citenamefont
  {Huang}(2023)}]{Yang:2023vsw}%
  \BibitemOpen
  \bibfield  {author} {\bibinfo {author} {\bibfnamefont {J.-C.}\ \bibnamefont
  {Yang}}\ and\ \bibinfo {author} {\bibfnamefont {X.-G.}\ \bibnamefont
  {Huang}},\ }\href@noop {} {\  (\bibinfo {year} {2023})},\ \Eprint
  {https://arxiv.org/abs/2307.05755} {arXiv:2307.05755 [hep-lat]} \BibitemShut
  {NoStop}%
\bibitem [{\citenamefont {Braguta}\ \emph {et~al.}(2024)\citenamefont
  {Braguta}, \citenamefont {Chernodub}, \citenamefont {Gershtein},\ and\
  \citenamefont {Roenko}}]{Braguta:2024zpi}%
  \BibitemOpen
  \bibfield  {author} {\bibinfo {author} {\bibfnamefont {V.~V.}\ \bibnamefont
  {Braguta}}, \bibinfo {author} {\bibfnamefont {M.~N.}\ \bibnamefont
  {Chernodub}}, \bibinfo {author} {\bibfnamefont {Y.~A.}\ \bibnamefont
  {Gershtein}},\ and\ \bibinfo {author} {\bibfnamefont {A.~A.}\ \bibnamefont
  {Roenko}},\ }\href@noop {} {\  (\bibinfo {year} {2024})},\ \Eprint
  {https://arxiv.org/abs/2411.15085} {arXiv:2411.15085 [hep-lat]} \BibitemShut
  {NoStop}%
\bibitem [{\citenamefont {Jiang}(2022)}]{Jiang2022}%
  \BibitemOpen
  \bibfield  {author} {\bibinfo {author} {\bibfnamefont {Y.}~\bibnamefont
  {Jiang}},\ }\href {https://doi.org/10.1140/epjc/s10052-022-10915-8}
  {\bibfield  {journal} {\bibinfo  {journal} {The European Physical Journal C}\
  }\textbf {\bibinfo {volume} {82}},\ \bibinfo {pages} {949} (\bibinfo {year}
  {2022})}\BibitemShut {NoStop}%
\bibitem [{\citenamefont {Nunes}\ \emph {et~al.}(2024)\citenamefont {Nunes},
  \citenamefont {Farias}, \citenamefont {Tavares},\ and\ \citenamefont
  {Tim\'oteo}}]{Nunes:2024hzy}%
  \BibitemOpen
  \bibfield  {author} {\bibinfo {author} {\bibfnamefont {R.~M.}\ \bibnamefont
  {Nunes}}, \bibinfo {author} {\bibfnamefont {R.~L.~S.}\ \bibnamefont
  {Farias}}, \bibinfo {author} {\bibfnamefont {W.~R.}\ \bibnamefont
  {Tavares}},\ and\ \bibinfo {author} {\bibfnamefont {V.~S.}\ \bibnamefont
  {Tim\'oteo}},\ }\href@noop {} {\  (\bibinfo {year} {2024})},\ \Eprint
  {https://arxiv.org/abs/2412.14541} {arXiv:2412.14541 [hep-ph]} \BibitemShut
  {NoStop}%
\bibitem [{\citenamefont {Jiang}\ and\ \citenamefont
  {Liao}(2016)}]{Jiang:2016wvv}%
  \BibitemOpen
  \bibfield  {author} {\bibinfo {author} {\bibfnamefont {Y.}~\bibnamefont
  {Jiang}}\ and\ \bibinfo {author} {\bibfnamefont {J.}~\bibnamefont {Liao}},\
  }\href {https://doi.org/10.1103/PhysRevLett.117.192302} {\bibfield  {journal}
  {\bibinfo  {journal} {Phys. Rev. Lett.}\ }\textbf {\bibinfo {volume} {117}},\
  \bibinfo {pages} {192302} (\bibinfo {year} {2016})},\ \Eprint
  {https://arxiv.org/abs/1606.03808} {arXiv:1606.03808 [hep-ph]} \BibitemShut
  {NoStop}%
\bibitem [{\citenamefont {Chernodub}\ and\ \citenamefont
  {Gongyo}(2017{\natexlab{a}})}]{Chernodub_2017}%
  \BibitemOpen
  \bibfield  {author} {\bibinfo {author} {\bibfnamefont {M.~N.}\ \bibnamefont
  {Chernodub}}\ and\ \bibinfo {author} {\bibfnamefont {S.}~\bibnamefont
  {Gongyo}},\ }\bibfield  {journal} {\bibinfo  {journal} {Journal of High
  Energy Physics}\ }\textbf {\bibinfo {volume} {2017}},\ \href
  {https://doi.org/10.1007/jhep01(2017)136} {10.1007/jhep01(2017)136} (\bibinfo
  {year} {2017}{\natexlab{a}})\BibitemShut {NoStop}%
\bibitem [{\citenamefont {Ebihara}\ \emph {et~al.}(2017)\citenamefont
  {Ebihara}, \citenamefont {Fukushima},\ and\ \citenamefont
  {Mameda}}]{Ebihara:2016fwa}%
  \BibitemOpen
  \bibfield  {author} {\bibinfo {author} {\bibfnamefont {S.}~\bibnamefont
  {Ebihara}}, \bibinfo {author} {\bibfnamefont {K.}~\bibnamefont {Fukushima}},\
  and\ \bibinfo {author} {\bibfnamefont {K.}~\bibnamefont {Mameda}},\ }\href
  {https://doi.org/10.1016/j.physletb.2016.11.010} {\bibfield  {journal}
  {\bibinfo  {journal} {Phys. Lett. B}\ }\textbf {\bibinfo {volume} {764}},\
  \bibinfo {pages} {94} (\bibinfo {year} {2017})},\ \Eprint
  {https://arxiv.org/abs/1608.00336} {arXiv:1608.00336 [hep-ph]} \BibitemShut
  {NoStop}%
\bibitem [{\citenamefont {Chernodub}\ and\ \citenamefont
  {Gongyo}(2017{\natexlab{b}})}]{Chernodub:2017ref}%
  \BibitemOpen
  \bibfield  {author} {\bibinfo {author} {\bibfnamefont {M.~N.}\ \bibnamefont
  {Chernodub}}\ and\ \bibinfo {author} {\bibfnamefont {S.}~\bibnamefont
  {Gongyo}},\ }\href {https://doi.org/10.1103/PhysRevD.95.096006} {\bibfield
  {journal} {\bibinfo  {journal} {Phys. Rev. D}\ }\textbf {\bibinfo {volume}
  {95}},\ \bibinfo {pages} {096006} (\bibinfo {year} {2017}{\natexlab{b}})},\
  \Eprint {https://arxiv.org/abs/1702.08266} {arXiv:1702.08266 [hep-th]}
  \BibitemShut {NoStop}%
\bibitem [{\citenamefont {Pollock}(2010)}]{Pollock:2010zz}%
  \BibitemOpen
  \bibfield  {author} {\bibinfo {author} {\bibfnamefont {M.~D.}\ \bibnamefont
  {Pollock}},\ }\href@noop {} {\bibfield  {journal} {\bibinfo  {journal} {Acta
  Phys. Polon. B}\ }\textbf {\bibinfo {volume} {41}},\ \bibinfo {pages} {1827}
  (\bibinfo {year} {2010})}\BibitemShut {NoStop}%
\bibitem [{\citenamefont {Yepez}(2011)}]{Yepez:2011bw}%
  \BibitemOpen
  \bibfield  {author} {\bibinfo {author} {\bibfnamefont {J.}~\bibnamefont
  {Yepez}},\ }\href@noop {} {\  (\bibinfo {year} {2011})},\ \Eprint
  {https://arxiv.org/abs/1106.2037} {arXiv:1106.2037 [gr-qc]} \BibitemShut
  {NoStop}%
\bibitem [{\citenamefont {Kapusta}\ \emph {et~al.}(2020)\citenamefont
  {Kapusta}, \citenamefont {Rrapaj},\ and\ \citenamefont
  {Rudaz}}]{Kapusta:2019sad}%
  \BibitemOpen
  \bibfield  {author} {\bibinfo {author} {\bibfnamefont {J.~I.}\ \bibnamefont
  {Kapusta}}, \bibinfo {author} {\bibfnamefont {E.}~\bibnamefont {Rrapaj}},\
  and\ \bibinfo {author} {\bibfnamefont {S.}~\bibnamefont {Rudaz}},\ }\href
  {https://doi.org/10.1103/PhysRevC.101.024907} {\bibfield  {journal} {\bibinfo
   {journal} {Phys. Rev. C}\ }\textbf {\bibinfo {volume} {101}},\ \bibinfo
  {pages} {024907} (\bibinfo {year} {2020})},\ \Eprint
  {https://arxiv.org/abs/1907.10750} {arXiv:1907.10750 [nucl-th]} \BibitemShut
  {NoStop}%
\bibitem [{\citenamefont {Brill}\ and\ \citenamefont
  {Wheeler}(1957)}]{RevModPhys.29.465}%
  \BibitemOpen
  \bibfield  {author} {\bibinfo {author} {\bibfnamefont {D.~R.}\ \bibnamefont
  {Brill}}\ and\ \bibinfo {author} {\bibfnamefont {J.~A.}\ \bibnamefont
  {Wheeler}},\ }\href {https://doi.org/10.1103/RevModPhys.29.465} {\bibfield
  {journal} {\bibinfo  {journal} {Rev. Mod. Phys.}\ }\textbf {\bibinfo {volume}
  {29}},\ \bibinfo {pages} {465} (\bibinfo {year} {1957})}\BibitemShut
  {NoStop}%
\bibitem [{\citenamefont {Padhan}\ \emph {et~al.}(2024)\citenamefont {Padhan},
  \citenamefont {Dwibedi}, \citenamefont {Chatterjee},\ and\ \citenamefont
  {Ghosh}}]{Padhan:2024edf}%
  \BibitemOpen
  \bibfield  {author} {\bibinfo {author} {\bibfnamefont {N.}~\bibnamefont
  {Padhan}}, \bibinfo {author} {\bibfnamefont {A.}~\bibnamefont {Dwibedi}},
  \bibinfo {author} {\bibfnamefont {A.}~\bibnamefont {Chatterjee}},\ and\
  \bibinfo {author} {\bibfnamefont {S.}~\bibnamefont {Ghosh}},\ }\href
  {https://doi.org/10.1103/PhysRevC.110.024904} {\bibfield  {journal} {\bibinfo
   {journal} {Phys. Rev. C}\ }\textbf {\bibinfo {volume} {110}},\ \bibinfo
  {pages} {024904} (\bibinfo {year} {2024})},\ \Eprint
  {https://arxiv.org/abs/2403.16647} {arXiv:2403.16647 [hep-ph]} \BibitemShut
  {NoStop}%
\bibitem [{\citenamefont {Chen}\ \emph {et~al.}(2021)\citenamefont {Chen},
  \citenamefont {Huang},\ and\ \citenamefont {Liao}}]{Chen2021}%
  \BibitemOpen
  \bibfield  {author} {\bibinfo {author} {\bibfnamefont {H.-L.}\ \bibnamefont
  {Chen}}, \bibinfo {author} {\bibfnamefont {X.-G.}\ \bibnamefont {Huang}},\
  and\ \bibinfo {author} {\bibfnamefont {J.}~\bibnamefont {Liao}},\ }\bibinfo
  {title} {{QCD} {P}hase {S}tructure {U}nder {R}otation},\ in\ \href
  {https://doi.org/10.1007/978-3-030-71427-7_11} {\emph {\bibinfo {booktitle}
  {Strongly Interacting Matter under Rotation}}},\ \bibinfo {editor} {edited
  by\ \bibinfo {editor} {\bibfnamefont {F.}~\bibnamefont {Becattini}}, \bibinfo
  {editor} {\bibfnamefont {J.}~\bibnamefont {Liao}},\ and\ \bibinfo {editor}
  {\bibfnamefont {M.}~\bibnamefont {Lisa}}}\ (\bibinfo  {publisher} {Springer
  International Publishing},\ \bibinfo {address} {Cham},\ \bibinfo {year}
  {2021})\ pp.\ \bibinfo {pages} {349--379}\BibitemShut {NoStop}%
\bibitem [{\citenamefont {Buballa}(2005)}]{Buballa:2003qv}%
  \BibitemOpen
  \bibfield  {author} {\bibinfo {author} {\bibfnamefont {M.}~\bibnamefont
  {Buballa}},\ }\href {https://doi.org/10.1016/j.physrep.2004.11.004}
  {\bibfield  {journal} {\bibinfo  {journal} {Phys. Rept.}\ }\textbf {\bibinfo
  {volume} {407}},\ \bibinfo {pages} {205} (\bibinfo {year} {2005})},\ \Eprint
  {https://arxiv.org/abs/hep-ph/0402234} {arXiv:hep-ph/0402234} \BibitemShut
  {NoStop}%
\bibitem [{\citenamefont {Ambruş}\ and\ \citenamefont
  {Winstanley}(2014)}]{Ambru__2014}%
  \BibitemOpen
  \bibfield  {author} {\bibinfo {author} {\bibfnamefont {V.~E.}\ \bibnamefont
  {Ambruş}}\ and\ \bibinfo {author} {\bibfnamefont {E.}~\bibnamefont
  {Winstanley}},\ }\href {https://doi.org/10.1016/j.physletb.2014.05.031}
  {\bibfield  {journal} {\bibinfo  {journal} {Physics Letters B}\ }\textbf
  {\bibinfo {volume} {734}},\ \bibinfo {pages} {296–301} (\bibinfo {year}
  {2014})}\BibitemShut {NoStop}%
\bibitem [{\citenamefont {Ambruş}\ and\ \citenamefont
  {Winstanley}(2016)}]{Ambru__2016}%
  \BibitemOpen
  \bibfield  {author} {\bibinfo {author} {\bibfnamefont {V.~E.}\ \bibnamefont
  {Ambruş}}\ and\ \bibinfo {author} {\bibfnamefont {E.}~\bibnamefont
  {Winstanley}},\ }\bibfield  {journal} {\bibinfo  {journal} {Physical Review
  D}\ }\textbf {\bibinfo {volume} {93}},\ \href
  {https://doi.org/10.1103/physrevd.93.104014} {10.1103/physrevd.93.104014}
  (\bibinfo {year} {2016})\BibitemShut {NoStop}%
\bibitem [{\citenamefont {Aung}\ \emph {et~al.}(2024)\citenamefont {Aung},
  \citenamefont {Dwibedi}, \citenamefont {Dey},\ and\ \citenamefont
  {Ghosh}}]{Aung:2023pjf}%
  \BibitemOpen
  \bibfield  {author} {\bibinfo {author} {\bibfnamefont {C.~W.}\ \bibnamefont
  {Aung}}, \bibinfo {author} {\bibfnamefont {A.}~\bibnamefont {Dwibedi}},
  \bibinfo {author} {\bibfnamefont {J.}~\bibnamefont {Dey}},\ and\ \bibinfo
  {author} {\bibfnamefont {S.}~\bibnamefont {Ghosh}},\ }\href
  {https://doi.org/10.1103/PhysRevC.109.034913} {\bibfield  {journal} {\bibinfo
   {journal} {Phys. Rev. C}\ }\textbf {\bibinfo {volume} {109}},\ \bibinfo
  {pages} {034913} (\bibinfo {year} {2024})},\ \Eprint
  {https://arxiv.org/abs/2303.16462} {arXiv:2303.16462 [nucl-th]} \BibitemShut
  {NoStop}%
\bibitem [{\citenamefont {Dwibedi}\ \emph {et~al.}(2024)\citenamefont
  {Dwibedi}, \citenamefont {Aung}, \citenamefont {Dey},\ and\ \citenamefont
  {Ghosh}}]{Dwibedi:2023akm}%
  \BibitemOpen
  \bibfield  {author} {\bibinfo {author} {\bibfnamefont {A.}~\bibnamefont
  {Dwibedi}}, \bibinfo {author} {\bibfnamefont {C.~W.}\ \bibnamefont {Aung}},
  \bibinfo {author} {\bibfnamefont {J.}~\bibnamefont {Dey}},\ and\ \bibinfo
  {author} {\bibfnamefont {S.}~\bibnamefont {Ghosh}},\ }\href
  {https://doi.org/10.1103/PhysRevC.109.034914} {\bibfield  {journal} {\bibinfo
   {journal} {Phys. Rev. C}\ }\textbf {\bibinfo {volume} {109}},\ \bibinfo
  {pages} {034914} (\bibinfo {year} {2024})},\ \Eprint
  {https://arxiv.org/abs/2305.10183} {arXiv:2305.10183 [nucl-th]} \BibitemShut
  {NoStop}%
\bibitem [{\citenamefont {Dwibedi}\ \emph {et~al.}(2025)\citenamefont
  {Dwibedi}, \citenamefont {Marattukalam}, \citenamefont {Padhan},
  \citenamefont {Chatterjee},\ and\ \citenamefont {Ghosh}}]{Dwibedi:2025bdd}%
  \BibitemOpen
  \bibfield  {author} {\bibinfo {author} {\bibfnamefont {A.}~\bibnamefont
  {Dwibedi}}, \bibinfo {author} {\bibfnamefont {D.~R.~J.}\ \bibnamefont
  {Marattukalam}}, \bibinfo {author} {\bibfnamefont {N.}~\bibnamefont
  {Padhan}}, \bibinfo {author} {\bibfnamefont {A.}~\bibnamefont {Chatterjee}},\
  and\ \bibinfo {author} {\bibfnamefont {S.}~\bibnamefont {Ghosh}},\
  }\href@noop {} {\  (\bibinfo {year} {2025})},\ \Eprint
  {https://arxiv.org/abs/2504.16049} {arXiv:2504.16049 [nucl-th]} \BibitemShut
  {NoStop}%
\bibitem [{\citenamefont {Romatschke}(2012)}]{Romatschke:2011qp}%
  \BibitemOpen
  \bibfield  {author} {\bibinfo {author} {\bibfnamefont {P.}~\bibnamefont
  {Romatschke}},\ }\href {https://doi.org/10.1103/PhysRevD.85.065012}
  {\bibfield  {journal} {\bibinfo  {journal} {Phys. Rev. D}\ }\textbf {\bibinfo
  {volume} {85}},\ \bibinfo {pages} {065012} (\bibinfo {year} {2012})},\
  \Eprint {https://arxiv.org/abs/1108.5561} {arXiv:1108.5561 [gr-qc]}
  \BibitemShut {NoStop}%
\bibitem [{\citenamefont {Tinti}\ \emph {et~al.}(2017)\citenamefont {Tinti},
  \citenamefont {Jaiswal},\ and\ \citenamefont {Ryblewski}}]{Tinti:2016bav}%
  \BibitemOpen
  \bibfield  {author} {\bibinfo {author} {\bibfnamefont {L.}~\bibnamefont
  {Tinti}}, \bibinfo {author} {\bibfnamefont {A.}~\bibnamefont {Jaiswal}},\
  and\ \bibinfo {author} {\bibfnamefont {R.}~\bibnamefont {Ryblewski}},\ }\href
  {https://doi.org/10.1103/PhysRevD.95.054007} {\bibfield  {journal} {\bibinfo
  {journal} {Phys. Rev. D}\ }\textbf {\bibinfo {volume} {95}},\ \bibinfo
  {pages} {054007} (\bibinfo {year} {2017})},\ \Eprint
  {https://arxiv.org/abs/1612.07329} {arXiv:1612.07329 [nucl-th]} \BibitemShut
  {NoStop}%
\bibitem [{\citenamefont {Albright}\ and\ \citenamefont
  {Kapusta}(2016)}]{Albright:2015fpa}%
  \BibitemOpen
  \bibfield  {author} {\bibinfo {author} {\bibfnamefont {M.}~\bibnamefont
  {Albright}}\ and\ \bibinfo {author} {\bibfnamefont {J.~I.}\ \bibnamefont
  {Kapusta}},\ }\href {https://doi.org/10.1103/PhysRevC.93.014903} {\bibfield
  {journal} {\bibinfo  {journal} {Phys. Rev. C}\ }\textbf {\bibinfo {volume}
  {93}},\ \bibinfo {pages} {014903} (\bibinfo {year} {2016})},\ \Eprint
  {https://arxiv.org/abs/1508.02696} {arXiv:1508.02696 [nucl-th]} \BibitemShut
  {NoStop}%
\bibitem [{\citenamefont {Nogarolli}\ \emph {et~al.}(2024)\citenamefont
  {Nogarolli}, \citenamefont {Denicol},\ and\ \citenamefont
  {Fraga}}]{Nogarolli:2024azy}%
  \BibitemOpen
  \bibfield  {author} {\bibinfo {author} {\bibfnamefont {P.}~\bibnamefont
  {Nogarolli}}, \bibinfo {author} {\bibfnamefont {G.~S.}\ \bibnamefont
  {Denicol}},\ and\ \bibinfo {author} {\bibfnamefont {E.~S.}\ \bibnamefont
  {Fraga}},\ }\href@noop {} {\  (\bibinfo {year} {2024})},\ \Eprint
  {https://arxiv.org/abs/2410.18791} {arXiv:2410.18791 [hep-ph]} \BibitemShut
  {NoStop}%
\bibitem [{\citenamefont {Rocha}\ and\ \citenamefont
  {Denicol}(2024)}]{Rocha:2024rce}%
  \BibitemOpen
  \bibfield  {author} {\bibinfo {author} {\bibfnamefont {G.~S.}\ \bibnamefont
  {Rocha}}\ and\ \bibinfo {author} {\bibfnamefont {G.~S.}\ \bibnamefont
  {Denicol}},\ }\href {https://doi.org/10.1103/PhysRevD.109.096011} {\bibfield
  {journal} {\bibinfo  {journal} {Phys. Rev. D}\ }\textbf {\bibinfo {volume}
  {109}},\ \bibinfo {pages} {096011} (\bibinfo {year} {2024})},\ \Eprint
  {https://arxiv.org/abs/2402.06996} {arXiv:2402.06996 [nucl-th]} \BibitemShut
  {NoStop}%
\bibitem [{\citenamefont {Anderson}\ and\ \citenamefont
  {Witting}(1974)}]{ANDERSON1974466}%
  \BibitemOpen
  \bibfield  {author} {\bibinfo {author} {\bibfnamefont {J.}~\bibnamefont
  {Anderson}}\ and\ \bibinfo {author} {\bibfnamefont {H.}~\bibnamefont
  {Witting}},\ }\href
  {https://doi.org/https://doi.org/10.1016/0031-8914(74)90355-3} {\bibfield
  {journal} {\bibinfo  {journal} {Physica}\ }\textbf {\bibinfo {volume} {74}},\
  \bibinfo {pages} {466} (\bibinfo {year} {1974})}\BibitemShut {NoStop}%
\bibitem [{\citenamefont {Jaiswal}(2013)}]{Jaiswal:2013vta}%
  \BibitemOpen
  \bibfield  {author} {\bibinfo {author} {\bibfnamefont {A.}~\bibnamefont
  {Jaiswal}},\ }\href {https://doi.org/10.1103/PhysRevC.88.021903} {\bibfield
  {journal} {\bibinfo  {journal} {Phys. Rev. C}\ }\textbf {\bibinfo {volume}
  {88}},\ \bibinfo {pages} {021903} (\bibinfo {year} {2013})},\ \Eprint
  {https://arxiv.org/abs/1305.3480} {arXiv:1305.3480 [nucl-th]} \BibitemShut
  {NoStop}%
\bibitem [{\citenamefont {Dey}\ \emph {et~al.}(2021)\citenamefont {Dey},
  \citenamefont {Satapathy}, \citenamefont {Murmu},\ and\ \citenamefont
  {Ghosh}}]{Dey:2019axu}%
  \BibitemOpen
  \bibfield  {author} {\bibinfo {author} {\bibfnamefont {J.}~\bibnamefont
  {Dey}}, \bibinfo {author} {\bibfnamefont {S.}~\bibnamefont {Satapathy}},
  \bibinfo {author} {\bibfnamefont {P.}~\bibnamefont {Murmu}},\ and\ \bibinfo
  {author} {\bibfnamefont {S.}~\bibnamefont {Ghosh}},\ }\href
  {https://doi.org/10.1007/s12043-021-02148-3} {\bibfield  {journal} {\bibinfo
  {journal} {Pramana}\ }\textbf {\bibinfo {volume} {95}},\ \bibinfo {pages}
  {125} (\bibinfo {year} {2021})},\ \Eprint {https://arxiv.org/abs/1907.11164}
  {arXiv:1907.11164 [hep-ph]} \BibitemShut {NoStop}%
\bibitem [{\citenamefont {Huang}\ \emph {et~al.}(2021)\citenamefont {Huang},
  \citenamefont {Liao}, \citenamefont {Wang},\ and\ \citenamefont
  {Xia}}]{XuGuangHuang:2020dtn}%
  \BibitemOpen
  \bibfield  {author} {\bibinfo {author} {\bibfnamefont {X.-G.}\ \bibnamefont
  {Huang}}, \bibinfo {author} {\bibfnamefont {J.}~\bibnamefont {Liao}},
  \bibinfo {author} {\bibfnamefont {Q.}~\bibnamefont {Wang}},\ and\ \bibinfo
  {author} {\bibfnamefont {X.-L.}\ \bibnamefont {Xia}},\ }\href
  {https://doi.org/10.1007/978-3-030-71427-7_9} {\bibfield  {journal} {\bibinfo
   {journal} {Lect. Notes Phys.}\ }\textbf {\bibinfo {volume} {987}},\ \bibinfo
  {pages} {281} (\bibinfo {year} {2021})},\ \Eprint
  {https://arxiv.org/abs/2010.08937} {arXiv:2010.08937 [nucl-th]} \BibitemShut
  {NoStop}%
\bibitem [{\citenamefont {Ivanov}\ and\ \citenamefont
  {Soldatov}(2017)}]{PhysRevC.95.054915}%
  \BibitemOpen
  \bibfield  {author} {\bibinfo {author} {\bibfnamefont {Y.~B.}\ \bibnamefont
  {Ivanov}}\ and\ \bibinfo {author} {\bibfnamefont {A.~A.}\ \bibnamefont
  {Soldatov}},\ }\href {https://doi.org/10.1103/PhysRevC.95.054915} {\bibfield
  {journal} {\bibinfo  {journal} {Phys. Rev. C}\ }\textbf {\bibinfo {volume}
  {95}},\ \bibinfo {pages} {054915} (\bibinfo {year} {2017})}\BibitemShut
  {NoStop}%
\bibitem [{\citenamefont {Deng}\ and\ \citenamefont
  {Huang}(2016)}]{XuGuangHuang2016}%
  \BibitemOpen
  \bibfield  {author} {\bibinfo {author} {\bibfnamefont {W.-T.}\ \bibnamefont
  {Deng}}\ and\ \bibinfo {author} {\bibfnamefont {X.-G.}\ \bibnamefont
  {Huang}},\ }\href {https://doi.org/10.1103/PhysRevC.93.064907} {\bibfield
  {journal} {\bibinfo  {journal} {Phys. Rev. C}\ }\textbf {\bibinfo {volume}
  {93}},\ \bibinfo {pages} {064907} (\bibinfo {year} {2016})},\ \Eprint
  {https://arxiv.org/abs/1603.06117} {arXiv:1603.06117 [nucl-th]} \BibitemShut
  {NoStop}%
\bibitem [{\citenamefont {Jiang}\ \emph {et~al.}(2016)\citenamefont {Jiang},
  \citenamefont {Lin},\ and\ \citenamefont {Liao}}]{Jiang2016}%
  \BibitemOpen
  \bibfield  {author} {\bibinfo {author} {\bibfnamefont {Y.}~\bibnamefont
  {Jiang}}, \bibinfo {author} {\bibfnamefont {Z.-W.}\ \bibnamefont {Lin}},\
  and\ \bibinfo {author} {\bibfnamefont {J.}~\bibnamefont {Liao}},\ }\href
  {https://doi.org/10.1103/PhysRevC.94.044910} {\bibfield  {journal} {\bibinfo
  {journal} {Phys. Rev. C}\ }\textbf {\bibinfo {volume} {94}},\ \bibinfo
  {pages} {044910} (\bibinfo {year} {2016})}\BibitemShut {NoStop}%
\bibitem [{\citenamefont {Cassing}\ \emph {et~al.}(2013)\citenamefont
  {Cassing}, \citenamefont {Linnyk}, \citenamefont {Steinert},\ and\
  \citenamefont {Ozvenchuk}}]{Cassing:2013iz}%
  \BibitemOpen
  \bibfield  {author} {\bibinfo {author} {\bibfnamefont {W.}~\bibnamefont
  {Cassing}}, \bibinfo {author} {\bibfnamefont {O.}~\bibnamefont {Linnyk}},
  \bibinfo {author} {\bibfnamefont {T.}~\bibnamefont {Steinert}},\ and\
  \bibinfo {author} {\bibfnamefont {V.}~\bibnamefont {Ozvenchuk}},\ }\href
  {https://doi.org/10.1103/PhysRevLett.110.182301} {\bibfield  {journal}
  {\bibinfo  {journal} {Phys. Rev. Lett.}\ }\textbf {\bibinfo {volume} {110}},\
  \bibinfo {pages} {182301} (\bibinfo {year} {2013})},\ \Eprint
  {https://arxiv.org/abs/1302.0906} {arXiv:1302.0906 [hep-ph]} \BibitemShut
  {NoStop}%
\bibitem [{\citenamefont {Fernandez-Fraile}\ and\ \citenamefont
  {Gomez~Nicola}(2009)}]{Fernandez-Fraile:2009eug}%
  \BibitemOpen
  \bibfield  {author} {\bibinfo {author} {\bibfnamefont {D.}~\bibnamefont
  {Fernandez-Fraile}}\ and\ \bibinfo {author} {\bibfnamefont {A.}~\bibnamefont
  {Gomez~Nicola}},\ }\href {https://doi.org/10.1140/epjc/s10052-009-0935-0}
  {\bibfield  {journal} {\bibinfo  {journal} {Eur. Phys. J. C}\ }\textbf
  {\bibinfo {volume} {62}},\ \bibinfo {pages} {37} (\bibinfo {year} {2009})},\
  \Eprint {https://arxiv.org/abs/0902.4829} {arXiv:0902.4829 [hep-ph]}
  \BibitemShut {NoStop}%
\bibitem [{\citenamefont {Greif}\ \emph {et~al.}(2014)\citenamefont {Greif},
  \citenamefont {Bouras}, \citenamefont {Greiner},\ and\ \citenamefont
  {Xu}}]{Greif2014}%
  \BibitemOpen
  \bibfield  {author} {\bibinfo {author} {\bibfnamefont {M.}~\bibnamefont
  {Greif}}, \bibinfo {author} {\bibfnamefont {I.}~\bibnamefont {Bouras}},
  \bibinfo {author} {\bibfnamefont {C.}~\bibnamefont {Greiner}},\ and\ \bibinfo
  {author} {\bibfnamefont {Z.}~\bibnamefont {Xu}},\ }\href
  {https://doi.org/10.1103/PhysRevD.90.094014} {\bibfield  {journal} {\bibinfo
  {journal} {Phys. Rev. D}\ }\textbf {\bibinfo {volume} {90}},\ \bibinfo
  {pages} {094014} (\bibinfo {year} {2014})}\BibitemShut {NoStop}%
\bibitem [{\citenamefont {Marty}\ \emph {et~al.}(2013)\citenamefont {Marty},
  \citenamefont {Bratkovskaya}, \citenamefont {Cassing}, \citenamefont
  {Aichelin},\ and\ \citenamefont {Berrehrah}}]{Marty:2013ita}%
  \BibitemOpen
  \bibfield  {author} {\bibinfo {author} {\bibfnamefont {R.}~\bibnamefont
  {Marty}}, \bibinfo {author} {\bibfnamefont {E.}~\bibnamefont {Bratkovskaya}},
  \bibinfo {author} {\bibfnamefont {W.}~\bibnamefont {Cassing}}, \bibinfo
  {author} {\bibfnamefont {J.}~\bibnamefont {Aichelin}},\ and\ \bibinfo
  {author} {\bibfnamefont {H.}~\bibnamefont {Berrehrah}},\ }\href
  {https://doi.org/10.1103/PhysRevC.88.045204} {\bibfield  {journal} {\bibinfo
  {journal} {Phys. Rev. C}\ }\textbf {\bibinfo {volume} {88}},\ \bibinfo
  {pages} {045204} (\bibinfo {year} {2013})},\ \Eprint
  {https://arxiv.org/abs/1305.7180} {arXiv:1305.7180 [hep-ph]} \BibitemShut
  {NoStop}%
\bibitem [{\citenamefont {Puglisi}\ \emph {et~al.}(2015)\citenamefont
  {Puglisi}, \citenamefont {Plumari},\ and\ \citenamefont
  {Greco}}]{Puglisi:2014pda}%
  \BibitemOpen
  \bibfield  {author} {\bibinfo {author} {\bibfnamefont {A.}~\bibnamefont
  {Puglisi}}, \bibinfo {author} {\bibfnamefont {S.}~\bibnamefont {Plumari}},\
  and\ \bibinfo {author} {\bibfnamefont {V.}~\bibnamefont {Greco}},\ }\href
  {https://doi.org/10.1016/j.physletb.2015.10.070} {\bibfield  {journal}
  {\bibinfo  {journal} {Phys. Lett. B}\ }\textbf {\bibinfo {volume} {751}},\
  \bibinfo {pages} {326} (\bibinfo {year} {2015})},\ \Eprint
  {https://arxiv.org/abs/1407.2559} {arXiv:1407.2559 [hep-ph]} \BibitemShut
  {NoStop}%
\bibitem [{\citenamefont {Amato}\ \emph {et~al.}(2013)\citenamefont {Amato},
  \citenamefont {Aarts}, \citenamefont {Allton}, \citenamefont {Giudice},
  \citenamefont {Hands},\ and\ \citenamefont {Skullerud}}]{Amato2013}%
  \BibitemOpen
  \bibfield  {author} {\bibinfo {author} {\bibfnamefont {A.}~\bibnamefont
  {Amato}}, \bibinfo {author} {\bibfnamefont {G.}~\bibnamefont {Aarts}},
  \bibinfo {author} {\bibfnamefont {C.}~\bibnamefont {Allton}}, \bibinfo
  {author} {\bibfnamefont {P.}~\bibnamefont {Giudice}}, \bibinfo {author}
  {\bibfnamefont {S.}~\bibnamefont {Hands}},\ and\ \bibinfo {author}
  {\bibfnamefont {J.-I.}\ \bibnamefont {Skullerud}},\ }\href
  {https://doi.org/10.1103/PhysRevLett.111.172001} {\bibfield  {journal}
  {\bibinfo  {journal} {Phys. Rev. Lett.}\ }\textbf {\bibinfo {volume} {111}},\
  \bibinfo {pages} {172001} (\bibinfo {year} {2013})}\BibitemShut {NoStop}%
\bibitem [{\citenamefont {Gorenstein}\ \emph {et~al.}(2008)\citenamefont
  {Gorenstein}, \citenamefont {Hauer},\ and\ \citenamefont
  {Moroz}}]{Gorenstein:2007mw}%
  \BibitemOpen
  \bibfield  {author} {\bibinfo {author} {\bibfnamefont {M.~I.}\ \bibnamefont
  {Gorenstein}}, \bibinfo {author} {\bibfnamefont {M.}~\bibnamefont {Hauer}},\
  and\ \bibinfo {author} {\bibfnamefont {O.~N.}\ \bibnamefont {Moroz}},\ }\href
  {https://doi.org/10.1103/PhysRevC.77.024911} {\bibfield  {journal} {\bibinfo
  {journal} {Phys. Rev. C}\ }\textbf {\bibinfo {volume} {77}},\ \bibinfo
  {pages} {024911} (\bibinfo {year} {2008})},\ \Eprint
  {https://arxiv.org/abs/0708.0137} {arXiv:0708.0137 [nucl-th]} \BibitemShut
  {NoStop}%
\bibitem [{\citenamefont {Itakura}\ \emph {et~al.}(2008)\citenamefont
  {Itakura}, \citenamefont {Morimatsu},\ and\ \citenamefont
  {Otomo}}]{Itakura:2007mx}%
  \BibitemOpen
  \bibfield  {author} {\bibinfo {author} {\bibfnamefont {K.}~\bibnamefont
  {Itakura}}, \bibinfo {author} {\bibfnamefont {O.}~\bibnamefont {Morimatsu}},\
  and\ \bibinfo {author} {\bibfnamefont {H.}~\bibnamefont {Otomo}},\ }\href
  {https://doi.org/10.1103/PhysRevD.77.014014} {\bibfield  {journal} {\bibinfo
  {journal} {Phys. Rev. D}\ }\textbf {\bibinfo {volume} {77}},\ \bibinfo
  {pages} {014014} (\bibinfo {year} {2008})},\ \Eprint
  {https://arxiv.org/abs/0711.1034} {arXiv:0711.1034 [hep-ph]} \BibitemShut
  {NoStop}%
\bibitem [{\citenamefont {Plumari}\ \emph {et~al.}(2012)\citenamefont
  {Plumari}, \citenamefont {Puglisi}, \citenamefont {Scardina},\ and\
  \citenamefont {Greco}}]{Plumari:2012ep}%
  \BibitemOpen
  \bibfield  {author} {\bibinfo {author} {\bibfnamefont {S.}~\bibnamefont
  {Plumari}}, \bibinfo {author} {\bibfnamefont {A.}~\bibnamefont {Puglisi}},
  \bibinfo {author} {\bibfnamefont {F.}~\bibnamefont {Scardina}},\ and\
  \bibinfo {author} {\bibfnamefont {V.}~\bibnamefont {Greco}},\ }\href
  {https://doi.org/10.1103/PhysRevC.86.054902} {\bibfield  {journal} {\bibinfo
  {journal} {Phys. Rev. C}\ }\textbf {\bibinfo {volume} {86}},\ \bibinfo
  {pages} {054902} (\bibinfo {year} {2012})},\ \Eprint
  {https://arxiv.org/abs/1208.0481} {arXiv:1208.0481 [nucl-th]} \BibitemShut
  {NoStop}%
\bibitem [{\citenamefont {Noronha-Hostler}\ \emph {et~al.}(2009)\citenamefont
  {Noronha-Hostler}, \citenamefont {Noronha},\ and\ \citenamefont
  {Greiner}}]{Noronha-Hostler:2008kkf}%
  \BibitemOpen
  \bibfield  {author} {\bibinfo {author} {\bibfnamefont {J.}~\bibnamefont
  {Noronha-Hostler}}, \bibinfo {author} {\bibfnamefont {J.}~\bibnamefont
  {Noronha}},\ and\ \bibinfo {author} {\bibfnamefont {C.}~\bibnamefont
  {Greiner}},\ }\href {https://doi.org/10.1103/PhysRevLett.103.172302}
  {\bibfield  {journal} {\bibinfo  {journal} {Phys. Rev. Lett.}\ }\textbf
  {\bibinfo {volume} {103}},\ \bibinfo {pages} {172302} (\bibinfo {year}
  {2009})},\ \Eprint {https://arxiv.org/abs/0811.1571} {arXiv:0811.1571
  [nucl-th]} \BibitemShut {NoStop}%
\bibitem [{Note1()}]{Note1}%
  \BibitemOpen
  \bibinfo {note} {In Refs.~\cite {Jiang2016}, $y-$ axis is chosen
  perpendicular to the reaction plane whereas we have defined $z-$ axis to be
  perpendicular to the reaction plane.}\BibitemShut {Stop}%
\end{thebibliography}%
\end{document}